\documentclass[preprint2,tighten]{aastex6}

\newcommand{\teff}{$\rm T_{\!\mbox{\scriptsize eff}}$}

\newcommand{\msun}{$M_\odot$}
\newcommand{\ebv}{$E(B-V)$}

\newcommand{\hi}{\mbox{\ion{H}{1}}}
\newcommand{\hii}{\mbox{\ion{H}{2}}}

\newcommand{\nii}{\mbox{[\ion{N}{2}]}}
\newcommand{\oiii}{\mbox{[\ion{O}{3}]}}
\newcommand{\oii}{\mbox{[\ion{O}{2}]}}
\newcommand{\sii}{\mbox{[\ion{S}{2}]}}
\newcommand{\siii}{\mbox{[\ion{S}{3}]}}
\newcommand{\oiir}{\mbox{\ion{O}{2}}}

\newcommand{\cel}{{\sc cel}}
\newcommand{\rl}{{\sc rl}}

\newcommand{\te}{$\mathrm {T_e}$}

\newcommand{\hgamma}{H$\gamma$}
\newcommand{\hbeta}{H$\beta$}
\newcommand{\halpha}{H$\alpha$}
\newcommand{\lin}{$\,\lambda$}
\newcommand{\llin}{$\,\lambda\lambda$}

\newcommand{\rtf}{$R_{25}$}

\newcommand{\logg}{$\log g$}
\newcommand{\loggf}{$\log g_F$}
\newcommand{\gf}{$g_F$}

\newcommand{\mbol}{$\rm m_{bol}$}
\newcommand{\fglr}{{\sc fglr}}
\newcommand{\trgb}{{\sc trgb}}

\newcommand{\oh}{12\,+\,log(O/H)}

\newcommand{\eo}{$\rm \epsilon(O)$}
\newcommand{\eosun}{$\rm \epsilon(O)_\odot$}

\newcommand{\rtwothree}{R23}

\newcommand{\vs}{vs.}
\newcommand{\eg}{e.g.}
\newcommand{\ie}{i.e.}
\newcommand{\hst}{{\em HST}}

\shorttitle{Young stars and ionized nebulae in M83}		% optional
\shortauthors{Bresolin et al.}				% optional

\begin{document}
%\title{Young stars and ionized nebulae in M83: comparing chemical abundances at high metallicity. \\Demise of the direct method?}
\title{Young stars and ionized nebulae in M83: \\comparing chemical abundances at high metallicity.}

\author{Fabio Bresolin and Rolf-Peter Kudritzki\altaffilmark{1}}
\affil{Institute for Astronomy, University of Hawaii \\
2680 Woodlawn Drive \\
Honolulu, HI 96822, USA}

\author{Miguel A. Urbaneja}
\affil{Institut f\"ur Astro- und Teilchenphysik, Universit\"at Innsbruck\\
Technikerstr. 25/8, 6020 Innsbruck, Austria}

\author{Wolfgang Gieren\altaffilmark{2}}
\affil{Departamento de Astronom\'ia, Universidad de Concepci\'on\\
Casilla 160-C, Concepci\'on, Chile}

\author{I-Ting Ho}
\affil{Institute for Astronomy, University of Hawaii \\
2680 Woodlawn Drive \\
Honolulu, HI 96822, USA}

\and

\author{Grzegorz Pietrzy\'nski\altaffilmark{3}}
\affil{Departamento de Astronom\'ia, Universidad de Concepci\'on\\
Casilla 160-C, Concepci\'on, Chile}

\altaffiltext{1}{University Observatory Munich, Scheinerstr. 1, D-81679 Munich, Germany}
\altaffiltext{2}{Millennium Institute of Astrophysics, Santiago, Chile}
\altaffiltext{3}{Nicolaus Copernicus Astronomical Center, Polish Academy of Sciences, ul. Bartycka 18, PL-00-716 Warszawa, Poland}

%==========================================================================================================

\begin{abstract}
We present spectra of 14 A-type supergiants in the metal-rich spiral galaxy M83. We derive stellar parameters and metallicities, and measure a spectroscopic distance modulus  
$\rm \mu = 28.47 \pm 0.10$ ($4.9 \pm 0.2$~Mpc), in agreement with other methods. We use the stellar characteristic metallicity of M83 and other systems to discuss a version of the galaxy mass-metallicity relation that is independent of the analysis of nebular emission lines and the associated systematic uncertainties. 
We reproduce the radial metallicity gradient of M83, which flattens at large radii, with a chemical evolution model, constraining gas inflow and outflow processes.
We carry out a comparative analysis of the metallicities we derive from the stellar spectra and published \hii\ region line fluxes, utilizing both the direct, \te-based method and different strong-line abundance diagnostics. 
The direct abundances are in relatively good agreement with the stellar metallicities, once 
we apply a modest correction to the nebular oxygen abundance due to depletion onto dust. Popular empirically calibrated strong-line diagnostics tend to provide nebular abundances that underestimate the stellar metallicities above the solar value by $\sim$0.2 dex. This result could be related to difficulties in selecting calibration samples at high metallicity. The O3N2 method calibrated by Pettini and Pagel gives the best agreement with our stellar metallicities.
We confirm that metal recombination lines yield nebular abundances that agree with the stellar abundances for high metallicity systems, but find evidence that in more metal-poor environments they tend to underestimate the stellar metallicities by a significant amount, opposite to the behavior of the direct method.
\end{abstract}


%==========================================================================================================
\section{Introduction} \label{sec:intro}
Measuring extragalactic chemical abundances is the key to deciphering a wide variety of physical and evolutionary processes occurring inside and between galaxies. For star-forming systems   
the investigation of the present-day abundances of the interstellar medium (ISM), photoionized by young massive stars, holds a prominent place in modern astronomy, laying the foundations of our understanding of the chemical evolution of the Universe. Regrettably, despite decades of observational and theoretical work, we still lack an absolute abundance scale, which is necessary for a complete and coherent picture of how the chemical elements are processed and moved around by galactic flows. 

The gas-phase metallicity, identified with the abundance of oxygen, the most common heavy element in the ISM, can be derived from forbidden, collisionally excited lines (\cel s) present in \hii\ region optical spectra.
Such evaluation depends critically on the knowledge of the physical conditions of the gas, in particular the electron temperature \te, because of the strong temperature sensitivity of the metal line emissivities (see the monograph by \citealt{Stasinska:2012} for a review).
In the so-called {\em direct}  method \te\ is obtained by the classical technique 
(\citealt{Menzel:1941}) that utilizes \cel s originating from transitions involving different energy levels of the same ions. The intensity ratio of the auroral \oiii\lin4363 to the nebular \oiii\llin4959,\,5007 lines can be  used to measure the temperature of the high-excitation zone, especially at low metallicities, where the weak auroral lines are more easily observed. The \nii\lin 5755/\lin6584 ratio is generally used for the low excitation zone.
Around the solar metallicity and above, as the increased gas cooling quenches the auroral lines, statistical methods, first introduced by \citet{Pagel:1979} and \citet{Alloin:1979}, relying on easily observed strong emission lines, complement or supplant altogether the use of the direct technique. As is well known, different strong-line diagnostics and calibration methodologies (\eg\ photoionization models \vs\ empirical \te\ derivations) yield substantial systematic offsets in the inferred gas metallicities (\citealt{Kennicutt:2003a, Moustakas:2010, Lopez-Sanchez:2012}), reaching values up to 0.7~dex (\citealt{Kewley:2008}). Methods calibrated from \te\ measurements tend to occupy the bottom of the abundance scale.

Small-scale departures from homogeneity in the
thermal (\citealt{Peimbert:1967}) and chemical abundance (\citealt{Tsamis:2003}) structure of photoionized nebulae, combined with the pronounced temperature dependence of \cel s, can bias the results obtained from the direct method to low values. A similar effect can also originate at high metallicity from large-scale temperature gradients (\citealt{Stasinska:2005}). 
Estimations of temperature fluctuations, parameterized by the mean square value $t^2$ (\citealt{Peimbert:1967}), indicate that optical \cel s underestimate the oxygen abundances by 0.2--0.3 dex (\citealt{Esteban:2004, Peimbert:2005}). This effect is usually regarded responsible for the systematic oxygen abundance offset of the same magnitude found between measurements from the direct method and the \oiir\ recombination lines (\rl s; \citealt{Peimbert:1993a, Garcia-Rojas:2007, Esteban:2009}). Discrepancies of comparable sizes are also obtained when the \te-based nebular abundances are compared to a theoretical analysis of the emission-line spectra
(\citealt{Blanc:2015, Vale-Asari:2016}). Such differences can also result from a non-thermal distribution of electron energies (\citealt{Nicholls:2012}). For the widely-used direct method the crux of the matter remains the fact that, with the presence of these effects, the \te\ values we measure from  optical \cel s  tend to overestimate the nebular temperatures, leading to systematically underestimated gas-phase metallicities. 

While the situation described above seems to spell doom for the direct method and its ability to produce correct nebular abundances, at least at high metallicity, there are various considerations that warrant further investigations involving \te-based abundances.
These include the existence of still poorly understood systematic uncertainties in photoionization models (\citealt{Blanc:2015}), the lack of clearly identified causes for temperature fluctuations in ionized nebulae (although several processes have been proposed, see \citealt{Peimbert:2006a})
and the possibility that recombination lines overestimate gas-phase oxygen abundances (\citealt{Ercolano:2007, Stasinska:2007a}). We also note that theoretical and observational considerations argue against the $\kappa$ electron velocity distribution (\citealt{Nicholls:2012, Nicholls:2013}) as a solution for the abundance discrepancies observed in photoionized nebulae (\citealt{Zhang:2016, Ferland:2016}).
\smallskip

In light of these difficulties, a complementary approach for the investigation of  present-day abundances in galaxies is the analysis of the surface chemical composition of early-type (OBA) stars, which in virtue of their young ages share the same initial chemical composition as their parent gas clouds and associated \hii\ regions. This is true in particular for elements, such as oxygen and iron, whose surface abundances are not significantly altered by evolutionary processes during most of the stellar lifetimes. Oxygen abundance comparisons between nearby B stars and \hii\ regions, as in the well-studied case of the Orion nebula (\citealt{Simon-Diaz:2011}), offer support for the nebular abundance scale defined by \rl s rather than \cel s. 
A salient consideration is that the systematic chemical abundance uncertainty for B- and A-type stars is on the order of 0.1~dex (\citealt{Przybilla:2006, Nieva:2012}), much smaller than for the analysis of nebular spectra.

For more than a decade our collaboration has focused on a project of stellar spectroscopy in nearby star-forming galaxies, with distances up to a few Mpc, selected for a long-term investigation of the distance scale (\citealt{Gieren:2005}), in order to measure the metal content of bright blue supergiant stars and their distances (see \citealt{Kudritzki:2016} and \citealt{Urbaneja:2016} for the most recent results and references). In comparing stellar with nebular abundances we found a varying degree of agreement, ranging from excellent (\eg\ in the case of NGC~300, \citealt{Bresolin:2009a}) to modest (with offsets $\sim$0.2 dex, as in the case of NGC~3109, \citealt{Hosek:2014}). There are also indications that especially for systems of relatively high metallicity, such as M31 (\citealt{Zurita:2012}) and the solar neighbourhood (\citealt{Simon-Diaz:2011, Garcia-Rojas:2014}), the \te\ method underestimates the stellar abundances.
\smallskip

In this paper we analyze new stellar spectra of blue supergiant stars obtained in the spiral galaxy M83 (NGC~5236), at a distance 
of 4.9~Mpc (\citealt[$1'' = 23.8$~pc]{Jacobs:2009}).
Our main motivation is to extend our stellar work to a galactic environment characterized by a high level of chemical enrichment, \ie~super-solar in the central regions, as already indicated by work on \hii\ regions  (\citealt{Bresolin:2002, Bresolin:2005}) and a single super star cluster (\citealt{Gazak:2014}). This is the metallicity regime where the systematic biases of the direct method should be more evident. We thus compare stellar and nebular metallicities using the \te\ method and a variety of strong line diagnostics, aiming to clarify how abundances inferred from the latter relate to the metallicities measured in young stars. In a nutshell, we find that \te-based abundances fare reasonably well in comparison with stellar metallicities across a wide range of abundances, but nevertheless that existing empirical calibrations of strong line methods can significantly underestimate the stellar abundances in the high-metallicity regime. We describe our observational material and the data reduction in Sect.~2, and the spectral analysis in Sect.~3. We derive a spectroscopic distance to M83 in Sect.~4. In Sect.~5 the stellar metallicities are used to discuss the mass-metallicity relation for nearby galaxies  and to compare with a variety of nebular abundance diagnostics. We develop a chemical evolution model to reproduce the radial metallicity distribution in M83 in Sect.~6. In our discussion in Sect.~7 we focus on the comparison of metallicities derived from the direct method, the blue supergiants and \rl s in a number of nearby galaxies, based on results published in the literature. In Sect.~8 we summarize our main conclusions.

%==========================================================================================================
\section{Observations and data reduction} \label{sec:data}
Blue supergiant candidates were identified from broadband {\em Hubble Space Telescope} (\hst) Wide Field Camera 3 (WFC3) images obtained as part of the Early Release Science Program (ID 11360; PI: R. O'Connell) and the General Observer Program 12513 (PI: W. Blair), presented by \citet{Blair:2014}. Our selection was based on magnitudes and colors (\eg\ F438W$-$F814W) that we deemed compatible with those expected for B and A supergiants at the distance of M83, moderately affected by interstellar extinction. We attempted to minimize the contaminating effects of \hii\ region line emission, by inspecting images in the F657N filter, as well as of close projected companion stars. The stellar magnitudes were measured with the {\sc Dolphot 2.0} package, a modified version of {\sc HST}phot (\citealt{Dolphin:2000}).

Spectra of our best candidates were obtained with the FOcal Reducer and low-dispersion Spectrograph FORS2 attached to the European Southern Observatory (ESO) Very Large Telescope (VLT) UT1 (Antu) telescope on Cerro Paranal. Two separate observing runs, on April 3-4, 2013 and April 22-23, 2015,  provided data for two separate, single pointings, each covering a $6\farcm8 \times 6\farcm8$ field of view. We used the Mask eXchange Unit (MXU) with 1 arc second slits to carry out multi-object spectroscopy of 42 and 24 targets for the two pointings, respectively (several slits were centered on 
\hii\ regions or late-type stars, and were not used for our analysis). The 600B grism yielded spectra with a resolution of 5\,\AA\ in the approximate wavelength range 3500--6000\,\AA.

We executed a series of 45 minute exposures throughout the duration of the observing runs, aiming at a good signal-to-noise ratio for our blue supergiant candidates. Degradation of the image quality at the higher airmasses and during spells of bad seeing or cloudy conditions limited our `effective' integration times, defined by the amount of useable spectra, to 12 hours (Apr 2013) and 7.5 hours (Apr 2015).

\begin{figure}
\center \includegraphics[width=1.01\columnwidth]{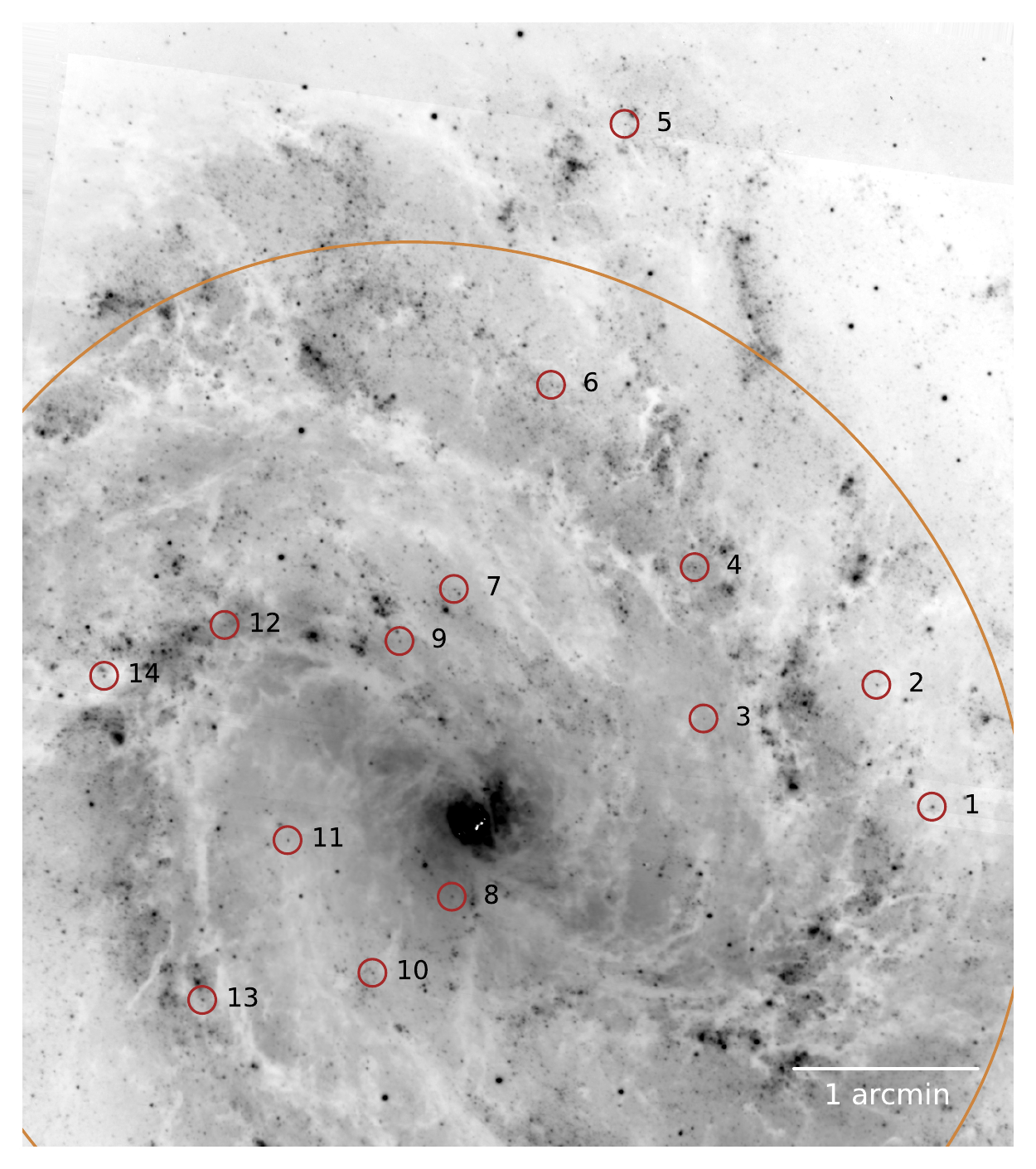}\medskip
\caption{The location of the blue supergiants analyzed in this work, shown on a $B$-band image taken with FORS2. The curve is drawn at a projected radius of 0.5\,\rtf. North is at the top, east on the left.}\label{fig:image}
\end{figure}

We reduced the data with the ESO Recipe Execution (EsoRex) tool, which provided us with wavelength-calibrated, two-dimensional spectra of each target. Standard {\sc iraf} routines were then used for the extraction and coaddition of the final spectra. During the examination of these final products we discarded a number of objects from further analysis due to poor signal-to-noise ratio, strong nebular contamination and the fact that some of the targets were of late spectral type (typically F), unsuited for our analysis techniques. We retained a total of 14 bona fide blue supergiants, nine from the April 2013 run and five from April 2015, having a signal-to-noise ratio in the continuum near \hgamma\ higher than 45. Their celestial coordinates and magnitudes, measured from the \hst\ images and converted to the Johnson-Cousins system, are summarized in Table~\ref{table:1}. The spectral types, ranging from B8 to A5, were estimated from a comparison with Galactic stellar templates. A map of the distribution of the stars in our final sample is shown in Fig.~\ref{fig:image}. The disk parameters used for the deprojection to galactocentric distances are given at the foot of Table~\ref{table:1}.

%==========================================================================================================

%==========================================================================================================
\floattable
%\onecolumngrid
\begin{deluxetable}{cCCccRcl}

\tablecolumns{8}
\tablewidth{0pt}
\tablecaption{Blue supergiants: observational properties.\label{table:1}}

\tablehead{
\colhead{ID}	     		&
\colhead{R.A.}	 			&
\colhead{DEC}	 			&
\colhead{$R$/\rtf}			&
\colhead{$V$}				&
\colhead{$B-V$}				&
\colhead{$V-I$}				&
\colhead{Spectral}\\[0.5mm]
\colhead{}       			&
\colhead{(J2000.0)}       	&
\colhead{(J2000.0)}       	&
\colhead{}					& 	
\colhead{}					& 
\colhead{}					& 		
\colhead{}					& 																			
\colhead{type}  } 				
\startdata
\\[-4mm]
01	&  13\, 36\, 49.11	&  -29\, 51\, 54.4	&  0.42	& 20.16 &  0.12   & \nodata	& B8    \\  %	B-15
02	&  13\, 36\, 50.50	&  -29\, 51\, 14.7	&  0.39	& 20.64 &  0.13   & \nodata	& A2-A5 \\  %	B-13
03	&  13\, 36\, 54.83	&  -29\, 51\, 25.7	&  0.24	& 21.48 &  0.12   & \nodata	& A2    \\  %	B-14
04	&  13\, 36\, 55.05	&  -29\, 50\, 36.4	&  0.31	& 19.90 &  0.15   & \nodata	& A2    \\  %	B-10
05	&  13\, 36\, 56.80	&  -29\, 48\, 11.9	&  0.64	& 21.43 & -0.01   & \nodata	& B9-A0 \\  %	B-01
06	&  13\, 36\, 58.64	&  -29\, 49\, 36.9	&  0.39	& 20.94 & \nodata & 0.42   	& A5    \\  %	A-11
07	&  13\, 37\, 01.08	&  -29\, 50\, 43.4	&  0.20	& 20.94 & \nodata & 0.39   	& A2    \\  %	A-17
08	&  13\, 37\, 01.16	&  -29\, 52\, 23.7	&  0.08	& 20.51 & \nodata & 0.38   	& A2-A5 \\  %	A-27
09	&  13\, 37\, 02.46	&  -29\, 51\, 00.4	&  0.15	& 20.95 & \nodata & 0.22   	& A0    \\  %	A-19
10	&  13\, 37\, 03.15	&  -29\, 52\, 48.5	&  0.17	& 20.67 & \nodata & 0.23   	& B8-B9 \\  %	A-29
11	&  13\, 37\, 05.27	&  -29\, 52\, 05.2	&  0.16	& 20.59 & \nodata & 0.30   	& A0    \\  %	A-26
12	&  13\, 37\, 06.84	&  -29\, 50\, 55.1	&  0.25	& 20.71 & \nodata & 0.40   	& A5    \\  %	A-18
13	&  13\, 37\, 07.41	&  -29\, 52\, 57.3	&  0.30	& 20.26 & \nodata & 0.48   	& B9-A0 \\  %	A-31
14	&  13\, 37\, 09.86	&  -29\, 51\, 11.7	&  0.33	& 20.38 & \nodata & 0.55   	& A2    \\[1mm]  %	A-21
\enddata
\tablecomments{$R$/\rtf\ calculated adopting the following disk geometry: i\,=\,24~deg, PA\,=\,45~deg (\citealt{Comte:1981}),
\rtf\,=\,6.44 arcmin (\citealt{de-Vaucouleurs:1991})~=~9.18~kpc.}
\end{deluxetable}
%\twocolumngrid

%==========================================================================================================
\floattable
\begin{deluxetable}{cRCCRCCR}

\tablecolumns{8}
\tablewidth{0pt}
\tablecaption{Blue supergiants: derived quantities.\label{table:2}}

\tablehead{
\colhead{ID}	     		&
\colhead{\teff}	 			&
\colhead{\logg}	 			&
\colhead{\loggf}			&
\colhead{[Z]}				&
\colhead{\mbol}				&
\colhead{\ebv}				&
\colhead{BC}\\[0.5mm]
\colhead{}       			&
\colhead{K}       	&
\colhead{cgs}       	&
\colhead{cgs}					& 	
\colhead{dex}					& 
\colhead{mag}					& 		
\colhead{mag}					& 																			
\colhead{mag}  } 				
\startdata
\\[-4mm]
01	&  11050^{+100}_{-100}	&  1.36	&  1.19^{+0.05}_{-0.05}	& 0.05\pm 0.07  & 19.14\pm0.057 & 0.15\pm  0.01	&  -0.52\pm  0.04 \\ % B-15
02	&   8470^{+200}_{-200}	&  1.13	&  1.42^{+0.13}_{-0.13}	& 0.19\pm 0.12	& 20.38\pm 0.08	& 0.08\pm  0.02	&   0.02\pm  0.05 \\ % B-13
03	&   9100^{+200}_{-250}	&  1.45	&  1.61^{+0.10}_{-0.12}	& 0.12\pm 0.10	& 21.00\pm 0.06	& 0.12\pm  0.01	&  -0.09\pm  0.05 \\ % B-14
04	&   9000^{+120}_{-120}	&  1.10	&  1.28^{+0.07}_{-0.07}	& 0.17\pm 0.05	& 19.45\pm 0.05	& 0.11\pm  0.01	&  -0.10\pm  0.04 \\ % B-10
05	&   9800^{+300}_{-300}	&  1.51	&  1.55^{+0.07}_{-0.08}	&-0.10\pm 0.12	& 21.16\pm 0.07	& 0.01\pm  0.01	&  -0.24\pm  0.06 \\ % B-01
06	&   8300^{+100}_{-100}	&  1.00	&  1.32^{+0.09}_{-0.09}	& 0.00\pm 0.06	& 20.13\pm 0.07	& 0.25\pm  0.02	&   0.03\pm  0.04 \\ % A-11
07	&   9800^{+200}_{-200}	&  1.31	&  1.35^{+0.07}_{-0.07}	& 0.21\pm 0.10	& 19.83\pm 0.06	& 0.27\pm  0.01	&  -0.23\pm  0.05 \\ % A-17
08	&   8500^{+100}_{-70 }	&  0.85	&  1.13^{+0.07}_{-0.06}	& 0.25\pm 0.06	& 19.85\pm 0.07	& 0.18\pm  0.01	&  -0.05\pm  0.05 \\ % A-27
09	&  11350^{+200}_{-200}	&  1.50	&  1.28^{+0.05}_{-0.05}	& 0.28\pm 0.08	& 19.81\pm 0.06	& 0.18\pm  0.01	&  -0.52\pm  0.05 \\ % A-19
10	&  11100^{+200}_{-200}	&  1.45	&  1.27^{+0.05}_{-0.05}	& 0.15\pm 0.07	& 19.59\pm 0.06	& 0.18\pm  0.01	&  -0.49\pm  0.05 \\ % A-29
11	&  10000^{+150}_{-150}	&  1.35	&  1.35^{+0.06}_{-0.06}	& 0.33\pm 0.05	& 19.63\pm 0.04	& 0.21\pm  0.01	&  -0.26\pm  0.04 \\ % A-26
12	&   7900^{+200}_{-200}	&  0.90	&  1.31^{+0.20}_{-0.24}	& 0.22\pm 0.20	& 20.05\pm 0.11	& 0.24\pm  0.02	&   0.12\pm  0.08 \\ % A-18
13	&   9775^{+250}_{-250}	&  1.20	&  1.24^{+0.07}_{-0.07}	& 0.23\pm 0.09	& 18.92\pm 0.08	& 0.33\pm  0.01	&  -0.25\pm  0.07 \\ % A-31
14	&   9000^{+300}_{-300}	&  1.00	&  1.18^{+0.09}_{-0.08}	& 0.15\pm 0.11	& 19.11\pm 0.10	& 0.34\pm  0.01	&  -0.13\pm  0.09 \\[1mm] % A-21
\enddata
\end{deluxetable}

%==========================================================================================================
\section{Spectral analysis}
For the quantitative analysis of the stellar spectra of our A-type stars we followed the procedure explained in detail by \citet{Kudritzki:2014} and \citet{Hosek:2014}. We present here only a brief outline, and refer to those papers for an in-depth discussion of the methodology we employ. The essence of the procedure  is represented by a comparison between the observed, normalized spectra of our blue supergiants to a grid of line-blanketed model spectra, described by \citet{Kudritzki:2008, Kudritzki:2012}, which include a non-LTE line formation treatment based on the work of \citet{Przybilla:2006}.

\begin{figure*}
\center \includegraphics[width=1.5\columnwidth]{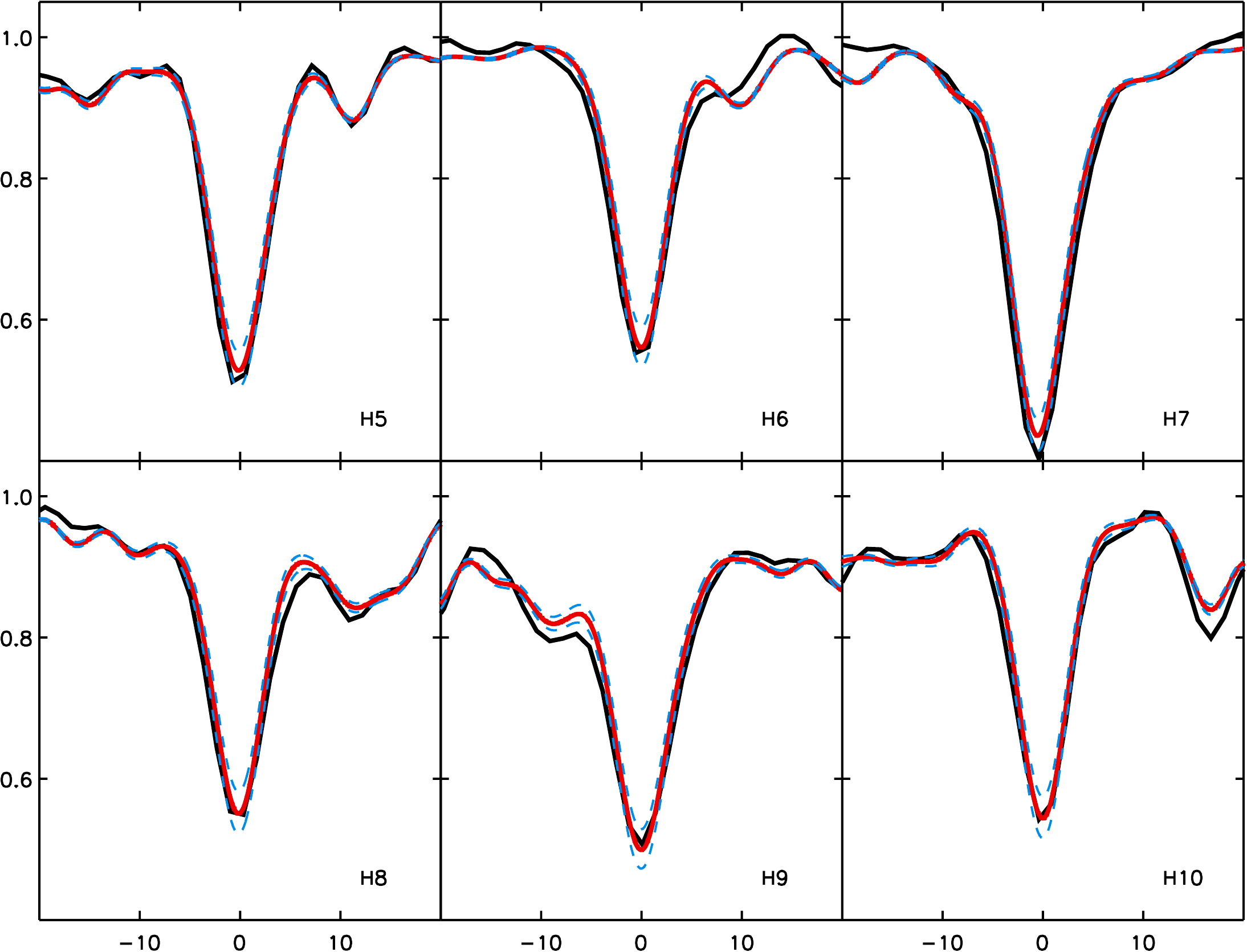}\medskip
\caption{Comparison between the observed high-order Balmer lines ($\rm H_5$ to $\rm H_{10}$) of star 02 (black profile) to the best-fitting synthetic spectrum (red profile). The dashed curves represent models deviating by $\pm0.05$ in \logg\ from the accepted solution.}\label{fig:balmer}
\end{figure*}

Firstly, the surface gravity (\logg) is derived as a function of stellar temperature (\teff), utilizing high-order Balmer lines (this choice minimizes the potential contamination by nebular emission). This step yields the \logg--\teff\ locus for each star. An example of line fitting is shown in Fig.~\ref{fig:balmer} for the mid-A star 02 (following the identification provided in Table~\ref{table:1}) and the Balmer spectral lines $\rm H_5$ to $\rm H_{10}$. The best-fitting model spectrum is drawn with the red continuous curve, flanked by models calculated using a $\pm0.05$ dex difference in \logg.

\begin{figure}
\center \includegraphics[width=1.01\columnwidth]{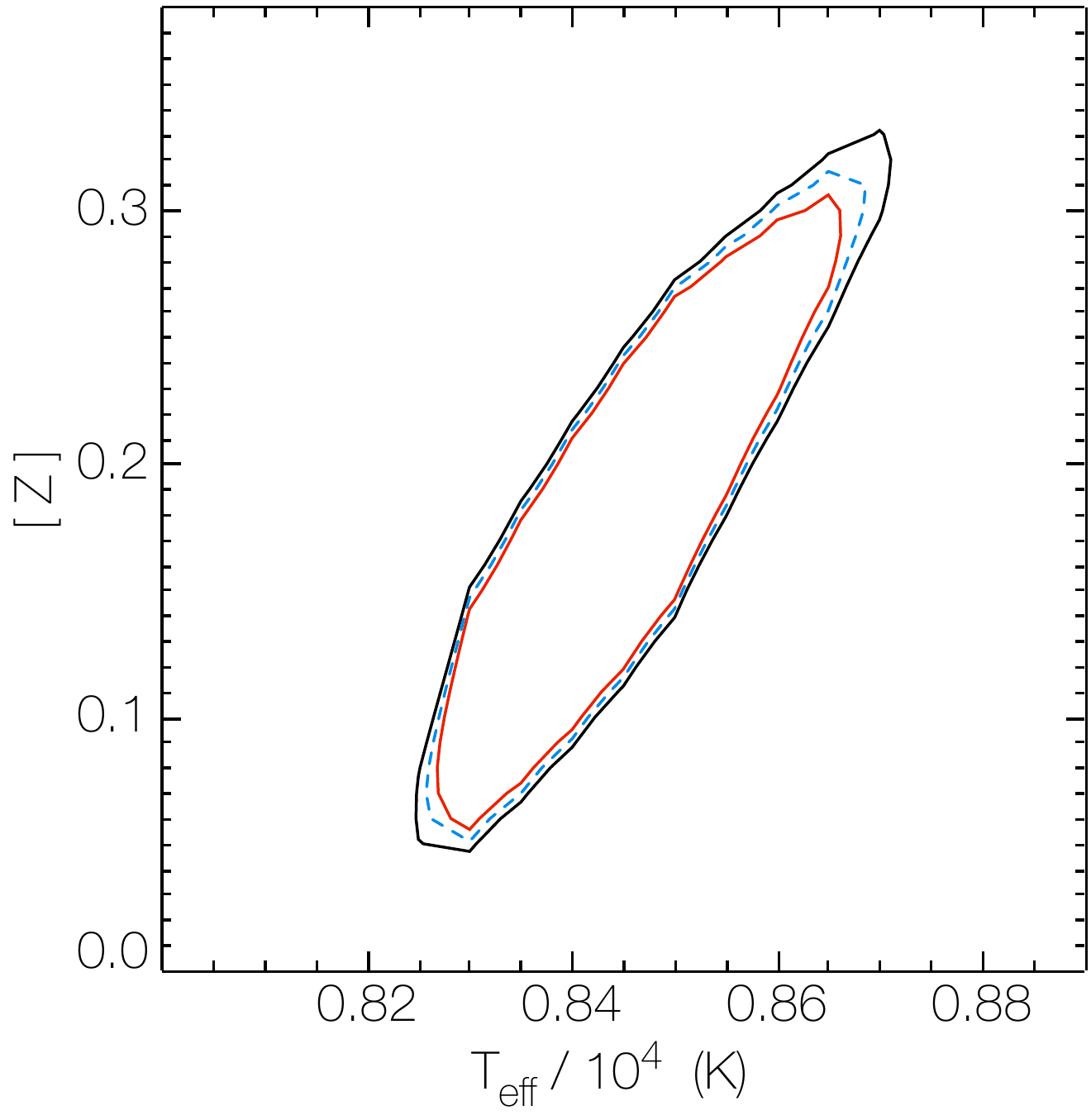}\medskip
\caption{$\Delta\chi^2$ isocontours in the metallicity-effective temperature plane for star 02. The curves are drawn for $\Delta\chi^2 = 3$ (inner, red), 6 (middle, blue) and 9 (outer, black). The adopted solution is [Z]~=~$0.19\pm0.12$, \teff~=~$8470\pm200$~K.}\label{fig:iso}
\end{figure}

\begin{figure*}
\center \includegraphics[width=1.5\columnwidth]{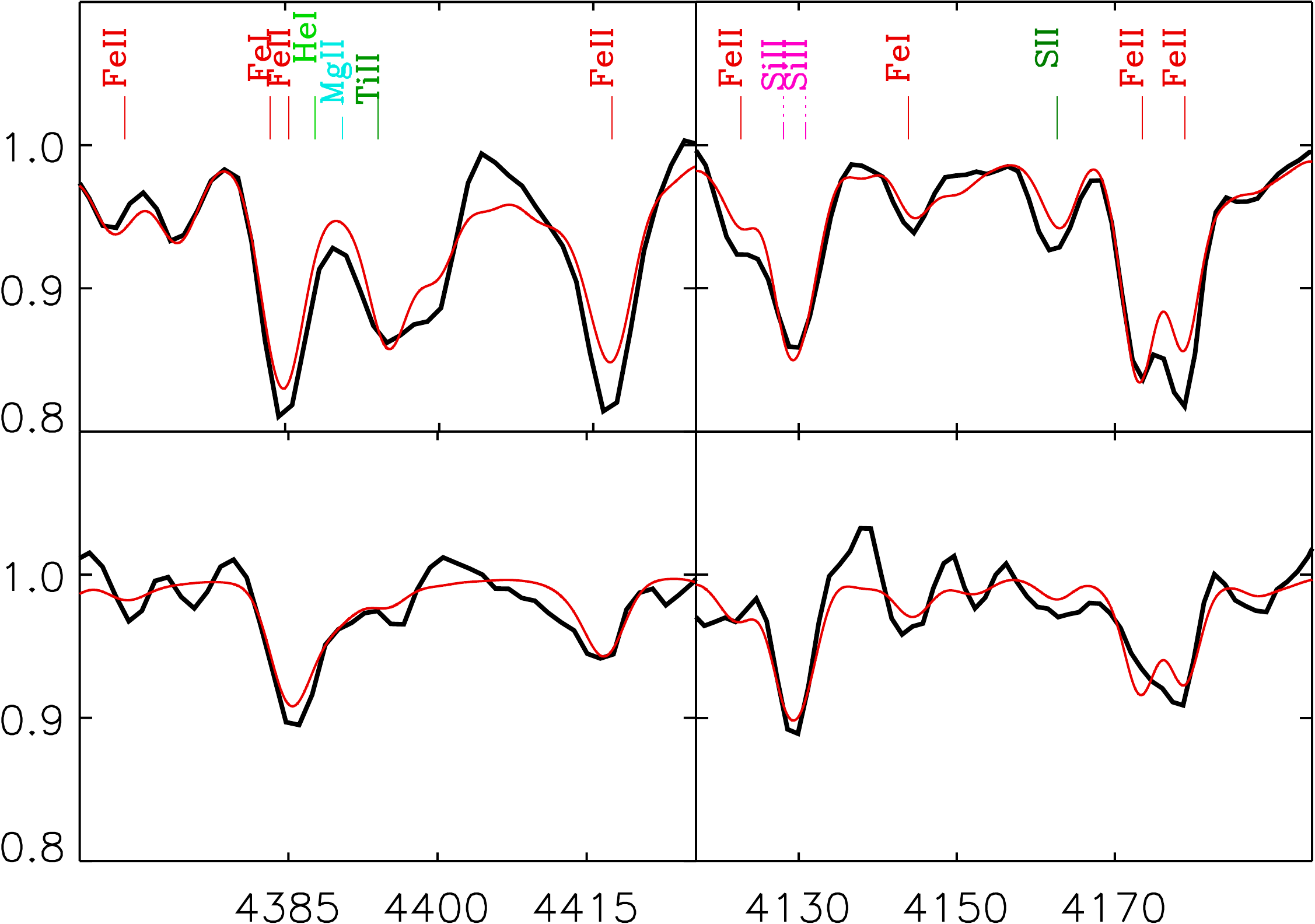}\medskip
\center \includegraphics[width=1.5\columnwidth]{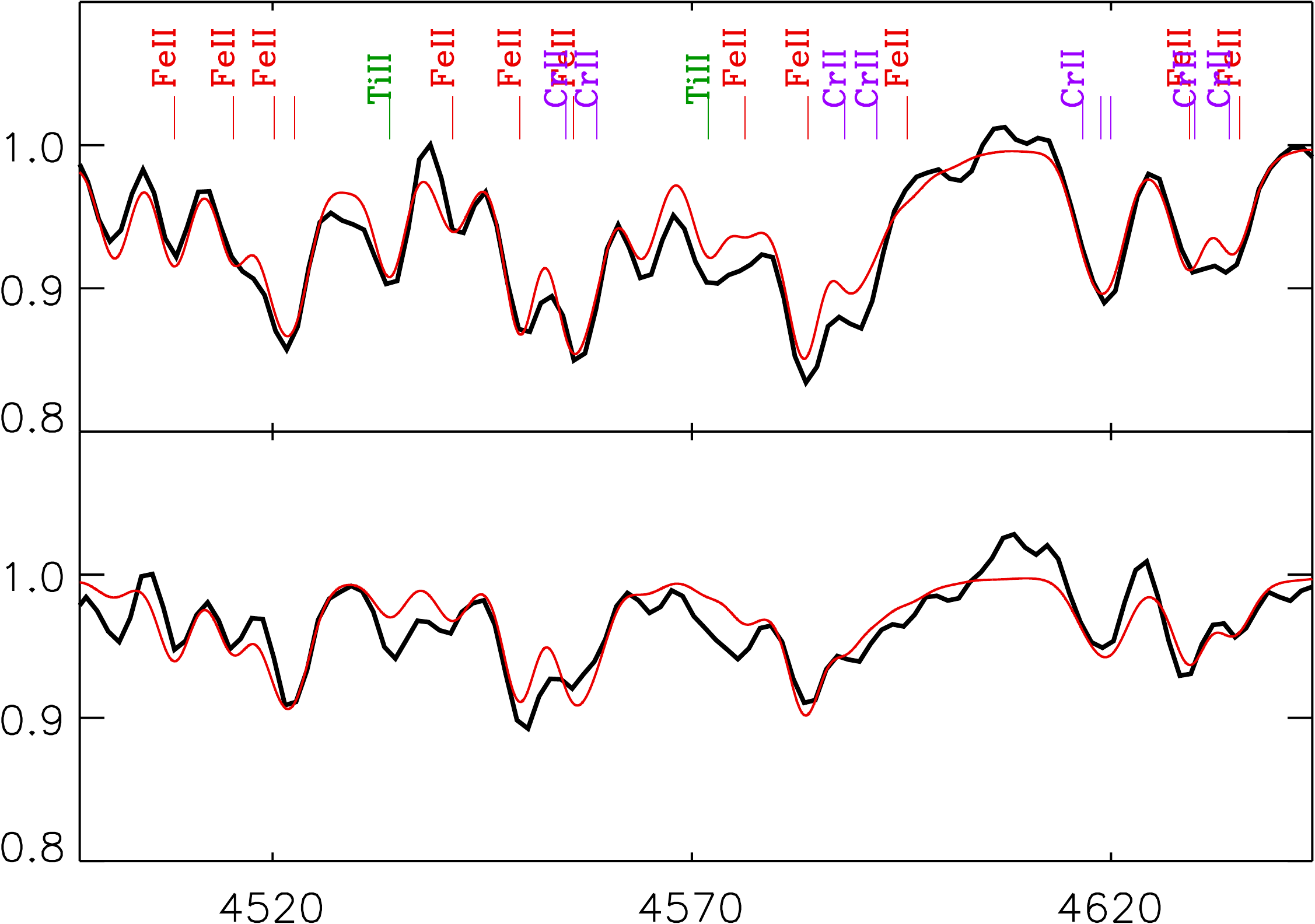}\medskip
\caption{Comparison of the observed normalized spectra (black) with synthetic spectra (red) calculated with the adopted stellar parameters. The two panels show different wavelength ranges, given in \AA\ along the $x$-axis. In each panel star 02 is at the top, star 07 at the bottom. The main atomic species responsible for the spectral lines calculated in the models are indicated.}\label{fig:metals}
\end{figure*}

Subsequently, we focus on the spectral lines produced by various chemical species present in 11 spectral windows in the wavelength range 3990--6000~\AA. 
We search for the best match between observed and synthetic spectra, varying the model metallicity and temperature values. The minimum $\chi^2$ value and the $\Delta\chi^2$ isocontours around this minimum allow us to define our adopted values for the metallicity and \teff\ and the associated uncertainties, as exemplified in Fig.~\ref{fig:iso}. The \logg--\teff\ relation derived in the first step finally provides us with the adopted surface gravity. The uncertainties in the stellar parameters are estimated from the $\chi^2$ procedure, adopting the $\Delta\chi^2 = 3$ isocontour as the 1-$\sigma$ uncertainty (\citealt{Hosek:2014}). In the case of \logg, the errors are derived from the uncertainty in \teff\ and a nominal error of 0.05~dex in the \logg--\teff\ relation. 
Fig.~\ref{fig:metals} illustrates the quality of the overall fit of the calculated metal line strengths to the observed spectra of stars 02 and 07 in three different wavelength ranges. We point out that the metallicity derived from our spectral synthesis procedure is a measure of the integrated chemical abundances of various elements, including both iron peak (Fe, Cr) and $\alpha$-elements (\eg\ Mg, Ca, Si, Ti). In our models we do not attempt to vary the $\alpha$/Fe abundance ratio. This aspect is relevant for the comparison with the nebular oxygen abundances carried out in Sect.~\ref{sec:hii}.

%==========================================================================================================
%==========================================================================================================

Finally, comparing the observed colors to the spectral energy distribution of the best-fitting stellar model provides us with the reddening \ebv, assuming a total-to-selective absorption ratio $R_V = A_V/E(B-V) = 3.3$.
  The selected model  also yields the bolometric correction (BC) used for the calculation of the bolometric magnitude (\mbol). The stellar parameters derived for our supergiant sample are summarized  in Table~\ref{table:2}, where we express
the metallicity with the standard notation $\rm [Z] = \log Z/Z_\odot$, but emphasizing that this is not strictly equivalent to the iron abundance. We also report the flux-weighted gravity \loggf = \logg $-4\log$\,(\teff/$10^4\,K$), that is used in Sect.~\ref{sec:fglr} to derive a spectroscopic distance to M83.

%==========================================================================================================

%==========================================================================================================

%==========================================================================================================

%==========================================================================================================

%==========================================================================================================
\subsection{Evolutionary status}
\begin{figure}
\center \includegraphics[width=1.01\columnwidth]{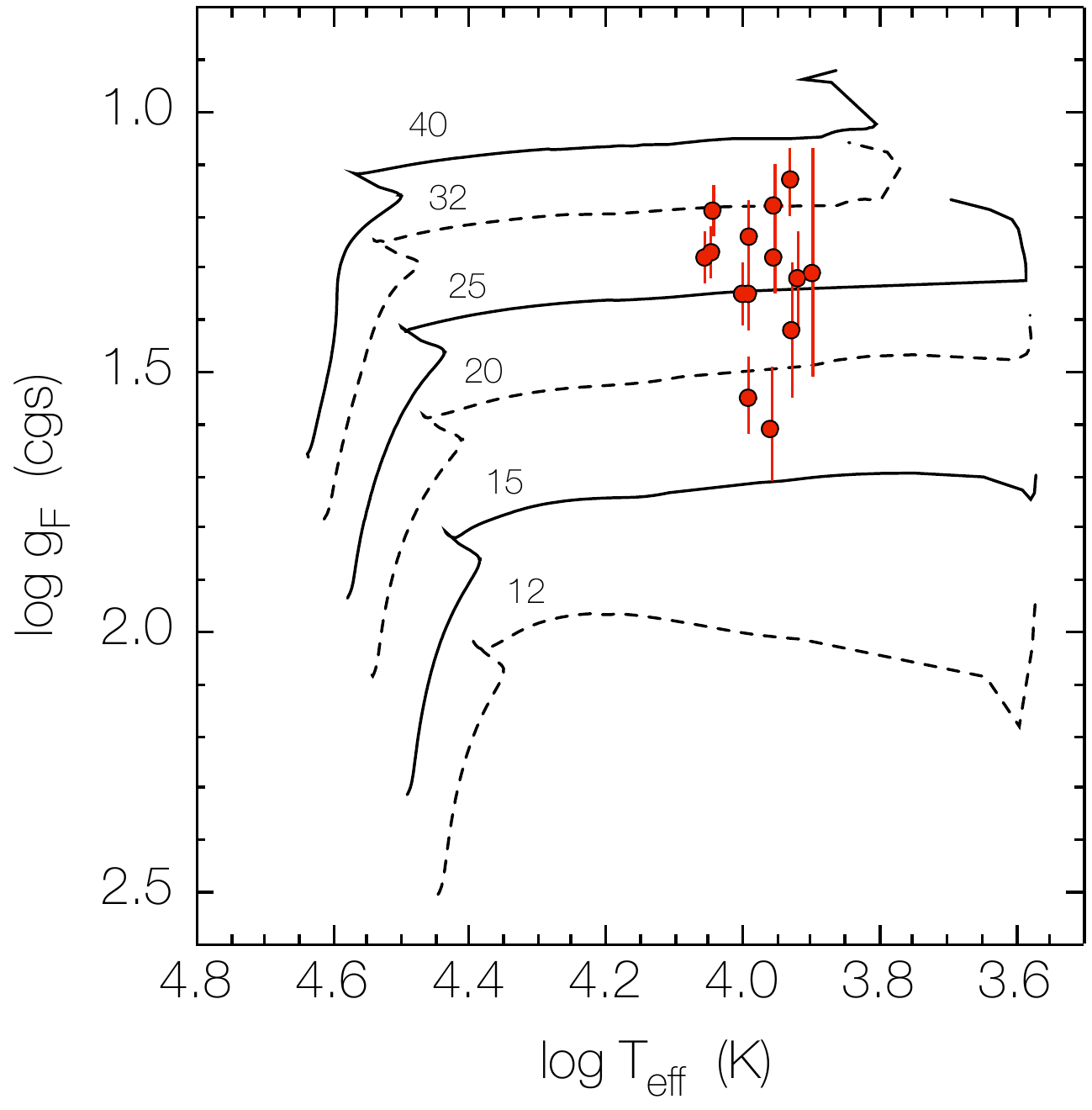}\medskip
\caption{Spectroscopic Hertzsprung-Russell diagram for our blue supergiant sample (dots with error bars). The curves represent stellar tracks (\citealt{Ekstrom:2012}) calculated for the initial masses indicated (in \msun).
}\label{fig:shrd}
\end{figure}

The distance-independent spectroscopic Hertzsprung-Russell diagram of our targets, relating the flux-weighted gravity \gf\ to the effective temperature \teff, following \citet{Langer:2014}, 
is displayed in Fig.~\ref{fig:shrd}. The stellar tracks for solar metallicity and including stellar rotation from \citet{Ekstrom:2012} are also shown, for masses between 12~\msun\ and 40~\msun. The diagram clearly illustrates the advanced evolutionary stage that pertains to these late-B--early-A supergiant stars. The tracks indicate that the supergiants in our sample  have initial main sequence masses comprised approximately between 15 and 32~\msun, in line with our previous investigations of blue supergiants in other nearby galaxies (\eg\ \citealt{Kudritzki:2012,Kudritzki:2013,Kudritzki:2014, Hosek:2014}).

%==========================================================================================================

%==========================================================================================================
\section{Spectroscopic distance}\label{sec:fglr}
As shown by \citet{Kudritzki:2003,Kudritzki:2008} the flux-weighted--luminosity relationship (\fglr), \ie\ the relation between \gf\ and \mbol, provides an independent method to measure extragalactic distances that utilizes medium-resolution spectra of blue supergiants. Among the advantages of this technique is the possibility of deriving both the reddening and the metallicity of the individual targets from the spectral analysis, allowing for tests of the dependence of other popular extragalactic distance indicators on, for example, chemical composition. In the series of papers from our group already mentioned (see \citealt{Kudritzki:2016} for a recent application) we have demonstrated how the \fglr\ provides distances that are generally in good agreement with the Cepheid period-luminosity (P-L) relation and the tip of the red giant branch (\trgb) methods.

\begin{figure}
\center \includegraphics[width=1.01\columnwidth]{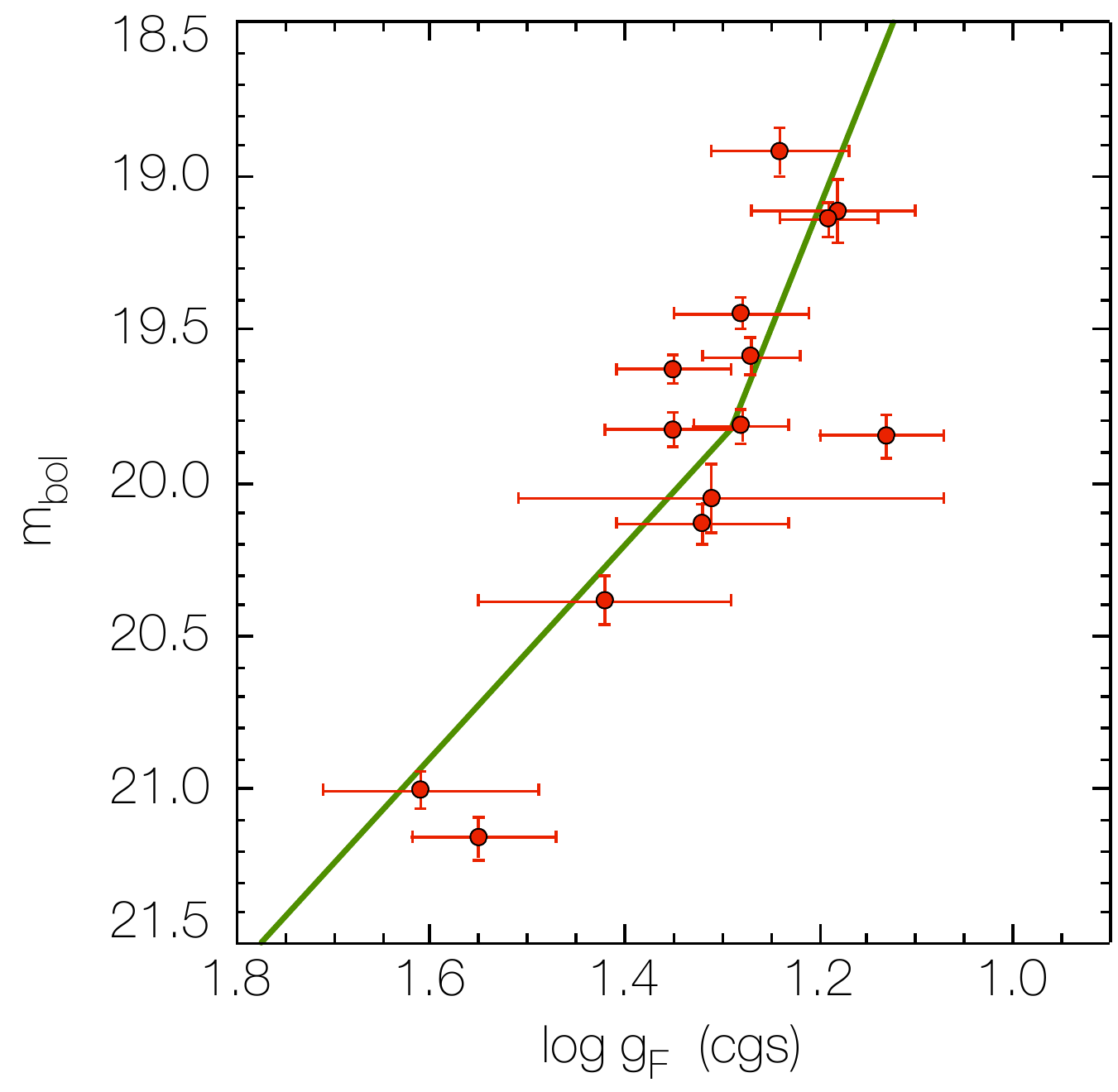}\medskip
\caption{The \fglr\ of our supergiants (red dots) and the fit to the calibrating relation by \citet[green line]{Urbaneja:2016}, used to derive the distance modulus $\rm \mu_{FGLR} = 28.47 \pm 0.10$.}\label{fig:fglr}
\end{figure}

To determine a \fglr-based distance to M83 we adopted the recent calibration of the technique published by \citet{Urbaneja:2016}, based on spectroscopic observations of 90 supergiants in the Large Magellanic Cloud, adopting a distance modulus to the LMC of $\rm \mu_{LMC} = 18.494$ from \citet{Pietrzynski:2013}.
The distance to M83 is obtained by  fitting their template \fglr\ to our individual stellar \loggf\ and \mbol\ values.
The result is shown in Fig.~\ref{fig:fglr}, where the steepening of the \fglr\ at high luminosities (small \loggf) found by \citet{Urbaneja:2016} is evident. From our procedure we obtain a distance modulus to M83 of $\rm \mu_{FGLR} = 28.47 \pm 0.10$ ($D = 4.9 \pm 0.2$~Mpc), where the error accounts for the observational uncertainties and those in the \fglr\ parameters.
Our independent determination of the distance modulus to M83 is in good agreement with the Cepheid P-L method, $\rm \mu_{P\textnormal{-}L} = 28.32 \pm 0.13$ (\citealt[28.27 if we adjust for our adopted LMC distance]{Saha:2006}), the \trgb\ method,  
$\rm \mu_{TRGB} = 28.45 \pm 0.04$ (\citealt{Jacobs:2009})
and the planetary nebula luminosity function, $\rm \mu_{PNLF} = 28.43 \pm 0.06$ (\citealt{Herrmann:2008}).

%==========================================================================================================

%==========================================================================================================

%==========================================================================================================
\section{Metallicity}\label{sec:metals}
In this section we take a detailed look at the metal content of blue supergiant stars in M83, one of our main motivations being the comparison with the chemical abundances of the ionized gas. First we  discuss M83 in the context of the mass-metallicity relation derived from stellar spectroscopy. The stellar mass of M83 has been determined following the procedure outlined by \citet{Kudritzki:2015}, using infrared surface photometry from the {\em Wide-field Infrared Survey Explorer} (WISE; \citealt{Wright:2010}) and the {\em Spitzer} Infrared Nearby Galaxies Survey (SINGS: \citealt{Kennicutt:2003a}). We obtained $\log M/M_\odot = 10.55$, adopting the spectroscopic distance from Sect.~\ref{sec:fglr}. The recent stellar mass analysis by \citet{Barnes:2014}, based on deep Spitzer Space Telescope imaging at 3.6~$\mu$m, yields $\log M/M_\odot = 10.72$, in good agreement with our result.
In the rest of the paper we will express the oxygen abundances with the notation \eo\,=\,\oh.

\subsection{Mass-metallicity relation}
The mass-metallicity relation (MZR; \citealt{Lequeux:1979, Tremonti:2004}) is an important diagnostic tool for galactic evolution studies, offering valuable insights into the star formation processes, the galactic wind outflows and the inflows that profoundly affect the chemical enrichment of the interstellar medium of star-forming galaxies (\citealt{Finlator:2008, Lilly:2013}), as well as into the redshift evolution of these mechanisms ({\citealt{Zahid:2014a, Sanders:2015}). While this relation is obtained almost exclusively from the emission-line analysis of star-forming galaxies, in our long-term project on extragalactic B- and A-type supergiants we have shown that a MZR can be defined for local galaxies using  metallicities measured from young supergiant stars.
Of course, the scope of this endeavor is not to compete with emission line studies, for the obvious reason that our sample size is orders of magnitude smaller. Our aim is to provide an independent look at this fundamental relation, based on a metallicity diagnostic for the young population that is completely distinct from the systematic uncertainties affecting the nebular abundances. In this context, this view of the MZR offers a first-order comparison between stellar and gaseous chemical abundances, adopting integrated metallicity values. In the following section we will consider a more detailed comparison, based on a spatially-resolved analysis of individual stars and \hii\ regions.

In Fig.~\ref{fig:mz} we show the MZR we have obtained from stellar chemical abundance studies, including all the galaxies in Table~10 of \citet{Hosek:2014}, with the addition of NGC~3621 (\citealt{Kudritzki:2014}), NGC~55 (\citealt{Kudritzki:2016}), and M83 (this study, red dot). 
The metallicity scale on the right axis is drawn adopting the solar oxygen abundance \eosun\,=\,8.69 from \citet{Asplund:2009}.
In the case of spiral galaxies, where the metallicity decreases with distance from the center, we adopt the characteristic value measured at 0.4~\rtf, based on the conclusion from \citet{Zaritsky:1994} and \citet{Moustakas:2006} that it coincides with the integrated metallicity. 

For comparison, in Fig.~\ref{fig:mz} we also include  the relations defined with the direct method  using galaxy stacks from the Sloan Digital Sky Survey (SDSS, Data Release~7, \citealt{Abazajian:2009}) by \citet[labeled as `SDSS']{Andrews:2013} and from a sample of dwarf irregular galaxies  by \citet[labeled as `dIrr']{Lee:2006}. The curves defined using SDSS %Data Release 4 (\citealt{Adelman-McCarthy:2006}) 
galaxies with four different strong-line nebular diagnostics, taken from \citet{Kewley:2008}, are also included. The four diagnostics are: N2, O3N2 (as calibrated by \citealt{Pettini:2004}), N2 (as calibrated combining photoionization models and empirical direct measurements by \citealt{Denicolo:2002}, labeled as D02) and \rtwothree\ (calibrated theoretically by \citealt{McGaugh:1991}, labeled as M91). \\

%==========================================================================================================

\begin{figure}
\center \includegraphics[width=1.01\columnwidth]{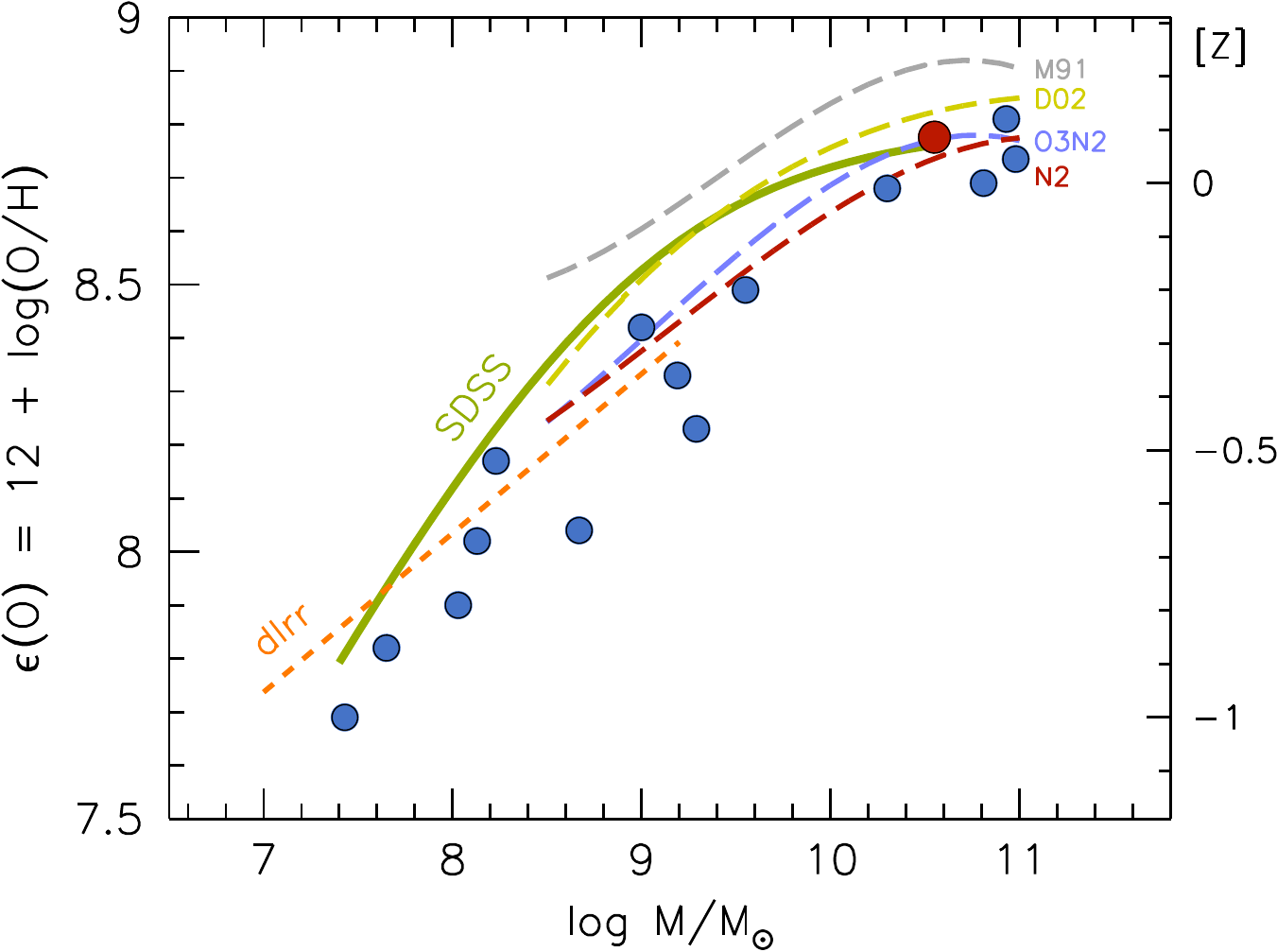}\medskip
\caption{The mass-metallicity relation determined from blue supergiant spectroscopy in 15 galaxies (dots). M83 is represented by the red dot. The lines show the relations defined by the direct method in galaxy stacks (\citealt[continuous green curve]{Andrews:2013}) and dwarf galaxies (\citealt[dashed orange  line]{Lee:2006}), and the relations defined from SDSS galaxies using four different strong-line diagnostics, taken from \citet[dashed lines]{Kewley:2008}.}\label{fig:mz}
\end{figure}

%==========================================================================================================

We highlight the following results from the comparison shown in Fig.~\ref{fig:mz}: 
\begin{itemize}

\item qualitatively the MZR based on stellar spectroscopy (`stellar' MZR) is similar to the relation obtained from nebular spectra, with a relatively modest scatter. There is not much evidence for a turnover or flattening of the MZR, except perhaps at high masses, which could be due to the unfavorable statistics.

\item in the high-mass regime ($\log M/M_\odot > 9.5$) the stellar MZR agrees significantly better with the MZR  derived using the N2 and O3N2 diagnostics rather than theoretical calibrations of abundance indicators such as \rtwothree\  (\citealt{McGaugh:1991}).

\item at intermediate masses ($8 < \log M/M_\odot < 9.5$) the stellar MZR deviates more significantly from the \te-based result by \citet{Andrews:2013} than at lower or higher masses. This suggests that the turnover in the stellar MZR occurs at higher masses, as also observed from the strong-line diagnostics. 
There is a better agreement with the curves obtained from both N2 and O3N2, and with the regression to the dwarf galaxy data by \citet{Lee:2006}.

\item at the lowest masses ($\log M/M_\odot < 9.5$) there is marginal agreement of the stellar MZR with \citet{Lee:2006} and \citet{Andrews:2013}.

\item the stellar MZR extends over a wide galactic stellar mass (3.5 dex), in fact extending to higher masses (and metallicities) than possible with the direct method, where the auroral lines become unmeasurable even in very high signal-to-noise SDSS spectral stacks.

\end{itemize}

We focus briefly on the comparison with \citet{Andrews:2013}, who have presented a recent determination of the MZR for star-forming galaxies in the local Universe, based on the stacking analysis of $\sim$200,000 SDSS galaxies.
Since blue supergiants tend to provide
{\em higher} metallicities  than the direct analysis of \hii\ regions in M83 (as shown in the next section), the fact that in Fig.~\ref{fig:mz} the stellar abundances appear offset to $\sim$0.2 {\em lower} abundances instead (at least for $\log M/M_\odot < 10$), might appear as inconsistent.
The more likely explanation is the effect of the star formation rate on the MZR calibrated by \citet{Andrews:2013}. 
%There are a few possible interpretations for this apparent paradox. If we defined our `characteristic' global metallicity at a smaller galactocentric radius than 0.4~\rtf\ for spiral galaxies, due to their radial abundance gradients we would measure higher metallicities. This effect can be important when comparing with SDSS, because of the potential aperture effect originating from the limited size on the sky of the SDSS fibers. We do note however that some of the discrepant points in Fig.~\ref{fig:mz} refer to galaxies with little or no radial abundance gradient.
%Another, more likely explanation, is provided by the effect of the star formation rate on the MZR calibrated by \citet{Andrews:2013}. 
These authors (see also \citealt{Brown:2016}) show how an increase of the 
SFR over the sample median shifts the MZR to lower O/H values. In fact, the average star-forming galaxy from the SDSS has a lower excitation than the \hii\ regions used to calibrate the
strong-line abundance indicators (\citealt{Pilyugin:2010a}), which produces the observed systematic metallicity offset.
This seems to be consistent with the fact that the dwarf irregular galaxies studied by \citet{Lee:2006}, in which the \te-based metallicity is generally measured from very few high-excitation \hii\ regions, define a MZR that is displaced to lower metallicities than the SDSS galaxies, except perhaps at the lowest masses, as shown in Fig.~\ref{fig:mz}.  

We conclude that the stellar MZR cannot be used to reliably infer which nebular abundance diagnostic yields metallicities that best match those of the supergiants. This is best done through a comparative, spatially-resolved analysis between supergiants and \hii\ regions, as done in the next section. 

%==========================================================================================================
\subsection{Comparison with gas abundances}\label{sec:hii}
Emission line fluxes of \hii\ regions in M83 have been taken from the following works: \citet{Bresolin:2002}, \citet{Bresolin:2005} and \citet{Bresolin:2009}. The former two focused on nebulae located in the inner disk ($R$ $<$ \rtf), while the latter studied the oxygen abundances in the extended, outer disk. We retain the full sample here, comprising 81 objects, even though the stars we studied are all situated at $R$ $<$ 0.64~\rtf. We calculated  strong-line abundances consistently, using a variety of methods, as described below. Direct abundances, based on the detection of auroral lines, are available for nine \hii\ regions, from \citet[5 objects in the inner disk]{Bresolin:2005} and \citet[4 objects in the outer disk]{Bresolin:2009}, including one in the nucleus of the galaxy, with a reported \eo\,=\,$8.94\pm0.09$ from \citet{Bresolin:2005}. As explained in Sect.~\ref{Sec:discussion}, we have redetermined the \te-based abundances, adopting a more recent set of atomic data, based on the references given in Table~5 of \citet{Bresolin:2009a}, and the \ion{O}{3} collision strengths from \citet{Palay:2012}. We now obtain 
\eo\,=\,$8.99\pm0.09$ for the central \hii\ region. We also note that for the inner disk regions the electron temperature has been measured based on the \nii\lin5755 and \siii\lin6312 auroral lines. The \oiii\lin4363 line has been detected in the outer disk \hii\ regions instead.

We also include in our comparison the young super star cluster (SSC) located near the center of the galaxy ($R$\,=\,0.06 \rtf), whose $J$-band spectral analysis has been presented by \citet{Gazak:2014}. These authors determined a metallicity [Z]\,=\,$0.28\pm0.14$, or 1.9$\times$ solar, which corresponds to \eo\,=\,$8.97\pm0.14$, confirming the high central metallicity indicated by the auroral lines detected in the nucleus. We revise this value to \eo\,=\,$8.88\pm0.14$, to account for the difference in the solar chemical composition adopted by us and the lower value adopted in the spectral synthesis based on MARCS models (\citealt{Gustafsson:2003}) carried out by \citet{Gazak:2014}.

In Fig.~\ref{fig:hii} we display metallicities, expressed as \eo, as a function of galactocentric distance, for the blue supergiants (star symbols), the SSC (cross symbol), \hii\ regions using the direct method (triangles) and \hii\ regions using different strong-line methods (circles). In each panel we vary the method used to derive the \hii\ region abundances, choosing, among the variety of   indicators available in the literature,  a representative set (both theoretically and empirically calibrated) that span the range of output abundance values.

Before looking in more detail at the strong-line methods, we make a couple of remarks. Near the center of M83 the stellar metallicities are in very good agreement with both the SSC analyzed by \citet{Gazak:2014} and the central \hii\ region auroral line analysis by \citet{Bresolin:2005}, with values \eo\,$\simeq$\,8.9--9.0 (1.6--2.0$\times$ solar). 
Of the remaining four \hii\ regions having auroral line detections and galactocentric radii in the range spanned by the blue supergiants, two have O/H values that are in good agreement with the stellar metallicities, while two have O/H values that are $\sim$0.3 dex below the stellar values. The radial decrease in metallicity in the inner disk appears to be steeper for the blue supergiants compared to the \hii\ regions, except perhaps for the case where the direct abundances are considered (but in this case the statistics is poor). It is certainly possible that this really depends on the limited range of galactocentric distances of the stars in our sample.

\subsubsection{Strong-line abundances}\label{sec:sl}
In this section we select a few representative strong-line diagnostics and compare the  chemical abundances they predict from the published line fluxes of \hii\ regions in M83, in order to
understand how they compare with the metallicities we have derived for the blue supergiants. It is not our scope to examine these diagnostics in detail, and we refer the interested reader to other discussions  in the literature (\eg~\citealt{Moustakas:2010, Lopez-Sanchez:2012, Blanc:2015}).

The strong-line abundance methods used in Fig.~\ref{fig:hii} are listed below (with the respective calibration paper given in brackets). We include two methods calibrated theoretically from photoionization model grids:

\bibpunct[\,]{(}{)}{,}{a}{}{;}		% changes the separation symbol in eg \citealt[=\,M91]{McGaugh:1991})

\renewcommand{\theenumi}{\alph{enumi}}
\begin{enumerate}

\item 
\rtwothree\ = (\oii\lin3727\,+\,\oiii\llin4959,\,5007)/\hbeta

(\citealt[=\,M91]{McGaugh:1991}) -- The calibration, in the analytical form given in \citet{Kuzio-de-Naray:2004}, accounts for changes in the ionization parameter, through the \oiii/\oii\ line ratio.

\item O2N2 = \nii\lin6584/\oii\lin3727\ (\citealt[=\,KD02]{Kewley:2002}) -- We adopt the calibration for a constant value of the ionization parameter $q = 2\times 10^7~ \rm cm\, s^{-1}$. This particular choice has little effect on the results, because of the weak effect of $q$ at high metallicities.

\end{enumerate}

%==========================================================================================================

\begin{figure*}
\center \includegraphics[width=2\columnwidth]{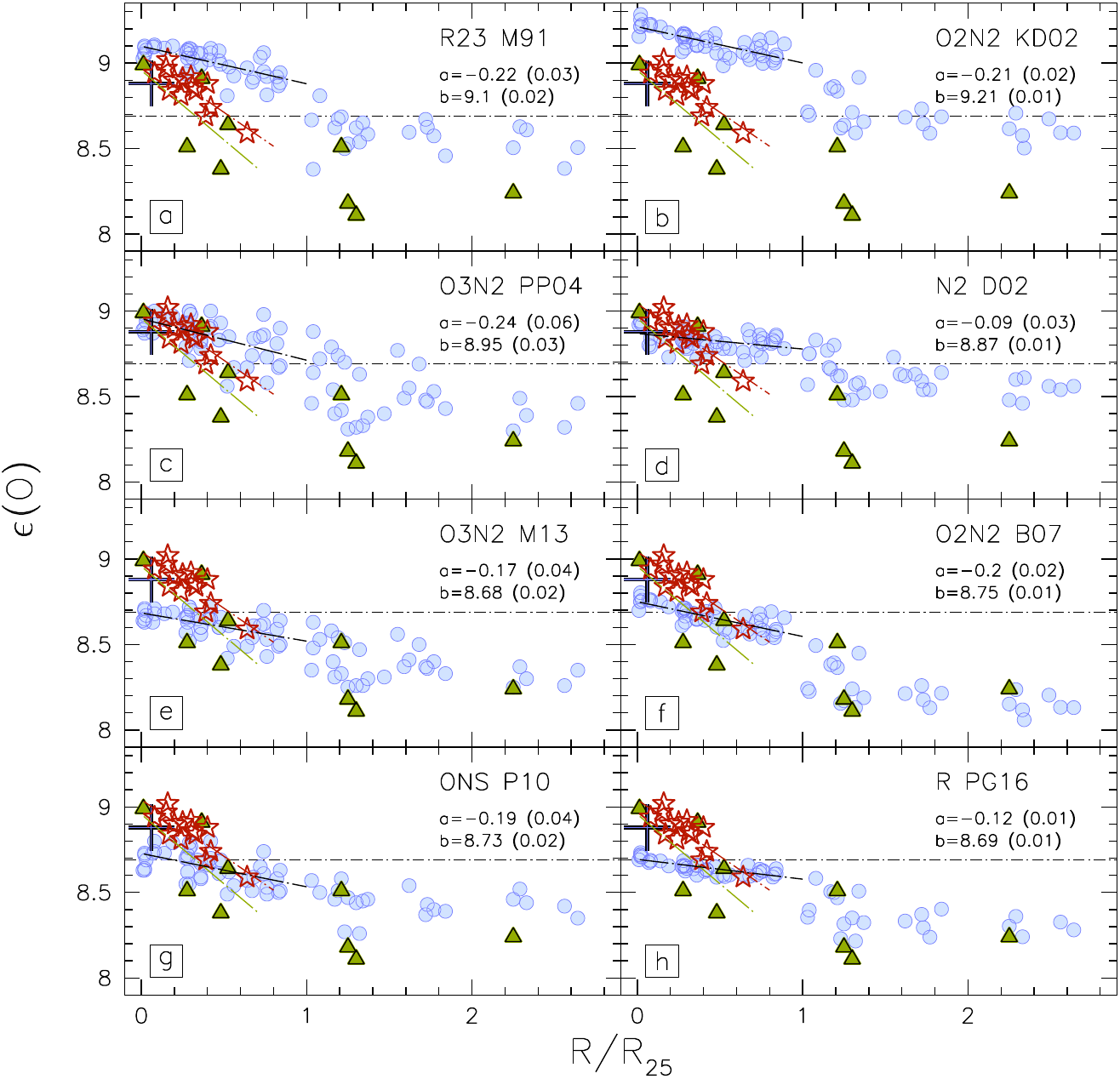}\medskip
\caption{Oxygen abundance vs. galactocentric distance, normalized to the isophotal radius, as determined from blue supergiants (red star symbols) and \hii\ regions in M83. The super star cluster studied by \citet{Gazak:2014} is shown by the green cross.
For \hii\ regions we plot direct (green triangles) and strong-line (blue circles) abundances. In each panel the nebular strong-line abundances were determined from a different diagnostic, as labeled (see text for more details). Linear regressions are shown for the different samples with dot-dashed lines. The slope ($a$) and the zero-point ($b$) of the regressions to the strong-line abundances are indicated in each panel, with errors in brackets. The horizontal line represents the adopted solar oxygen abundance.}\label{fig:hii}
\end{figure*}

%==========================================================================================================

Our stellar abundances lie 0.2--0.3 dex below the nebular metallicities calculated from both these diagnostics, except in the very central regions of M83, where the stellar and \rtwothree\ 
abundances converge.\\

The next two diagnostics were calibrated from a combination of theoretical models (at high metallicities) and empirical, \te-based abundances (at lower metallicities):

\begin{enumerate}
\setcounter{enumi}{2}

\item O3N2 = \oiii\lin5007/\hbeta\ $\cdot$ \halpha/\nii\lin6584\ (\citealt[=\,PP04]{Pettini:2004}) -- The calibration is based on 131 extragalactic \hii\ regions with direct abundances, supplemented with six nebulae whose oxygen abundances were determined with photoionization models. Four of these effectively shape the calibration at high metallicity, near \eo~$\simeq$~9.0.

\item N2 = \nii/\halpha\ (\citealt[=\,D02]{Denicolo:2002}) -- This is also a hybrid calibration, composed of \oiii\lin4363-based abundances and model results at higher metallicities. This diagnostic tends to saturate at metallicities above solar. This behavior can be associated to the limited O/H dynamic range observed in Fig.~\ref{fig:hii} with respect to the previous methods. The effect is even stronger considering the calibrations  by \citet{Pettini:2004} and \citet{Marino:2013} (not shown). 

\end{enumerate}

Using these two diagnostics yields the best agreement between nebular and stellar metallicities in M83, a somewhat surprising result when considering the more recent or updated abundance diagnostics presented below.
On the other hand, the agreement with the direct method abundances, shown by the green triangles, is overall rather poor. This is also unexpected, since both these calibrations are tied, except at high metallicities, to measurements of the auroral lines.\\

The remaining methods we consider are based on empirical calibrations, \ie\ on samples of extragalactic \hii\ regions where direct measurements of their oxygen abundances are available.

\begin{enumerate}
\setcounter{enumi}{4}

\item O3N2 (\citealt[=\,M13]{Marino:2013}) -- The recalibration of the O3N2 and N2 diagnostics by these authors is based on a compilation of direct abundances comprising 603 \hii\ regions. The result is a weaker dependence of these indices on metallicity compared to \citet{Pettini:2004}, producing the shallow abundance gradient in M83 seen in Fig.~\ref{fig:hii} (N2 yields a similar result). The stellar and nebular metallicities progressively diverge from each other with decreasing galactocentric radius, reaching a difference of $\sim$0.3 dex near the galaxy center.

\item O2N2 (\citealt[=\,B07]{Bresolin:2007}) -- About 140 direct abundance determinations were used in this case. This index seems to provide the best fit to the auroral line-based abundances among the indicators shown in Fig.~\ref{fig:hii}, albeit not perfect. The behavior of the abundance gradient at small galactocentric distance resembles what is seen in the case of O3N2 (M13), with a central discrepancy relative to the supergiants of $\sim$0.25 dex. %As already stated earlier, the auroral line measurements in M83 shown by the green triangles suggest an inner disk gradient that is steeper than can be inferred from any of the strong-line methods.

\item ONS (\citealt[=\,P10]{Pilyugin:2010}) -- This diagnostic, making use of the strengths of the \oii\lin3727, \oiii\llin4959,\,5007, \nii\llin6548,\,6584 and \sii\llin6717,\,6731 emission lines, provides results in the inner disk that are similar to the previous two. The comparison with the outer disk direct abundances is rather poor.

\item $R$ (\citealt[=\,PG16]{Pilyugin:2016}) -- The $R$ calibration makes use of the same lines 
as the ONS method, except for the exclusion of the \sii\ lines, and is based on a sample of 313 reference \hii\ regions of the `counterpart' method by the same authors (\citealt{Pilyugin:2012a}). Again, the overall outcome is comparable to the previous examples shown in Fig.~\ref{fig:hii}, the main difference being the extremely small abundance scatter obtained in the inner disk. We also tested the $S$ calibration (using \sii\  in place of \nii) from the same authors, and found results that are consistent with the $R$ calibration.

\end{enumerate}
\bigskip

\bibpunct[, ]{(}{)}{,}{a}{}{;}		% go back to default punctuation for natbib

Each panel of Fig.~\ref{fig:hii} reports the values of the slope and the zero point, with their  errors in brackets, of a linear regression of the form \eo\,=\,a ($R/$\rtf)\ + b
to the data points corresponding to the adopted strong-line indicator, for $< R/$\rtf. Our choice of the upper limit of the galactocentric distance range is somewhat arbitrary, but is necessary in order to exclude the outer disk \hii\ regions, which follow a flat radial abundance distribution (\citealt{Bresolin:2009}).
Black dot-dashed lines visualize the calculated regressions.  Each plot also shows linear regressions to the blue supergiant metallicities (red line), for which we obtain

\begin{equation}
\rm
\epsilon(O) = -0.66~ (\pm 0.13)~ {\it R/R}_{25}  + 9.04~ (\pm 0.04)
\end{equation}

and to the \hii\ region direct abundances (green line, only for $< R/$\rtf):

\begin{equation}
\rm
\epsilon(O) = -0.81~ (\pm 0.57)~ {\it R/R}_{25}  + 8.96~ (\pm 0.21)
\end{equation}

Keeping in mind the uncertainties due to the limited radial coverage of both the blue supergiants and the \hii\ regions with available \te-based abundances, these regressions show that {\em all} the strong-line indicators we included in Fig.~\ref{fig:hii} produce \hii\ region abundance gradients that are significantly shallower than either the blue supergiants or the direct method. We stress that this can be due to the small number of objects considered.
Blue supergiants and \hii\ regions with \te-based abundances have consistent slopes within the (large) uncertainties, with the stars offset by $\sim$0.1 dex to higher metallicity. 
On the other hand, panel $c$ of Fig.~\ref{fig:hii} also suggests that the radial metallicity  distribution and the scatter of our stellar targets are not dissimilar from what can be obtained from \hii\ region data and the O3N2 PP04 method. The measurement of blue supergiant abundances at larger galactocentric distances would be necessary to draw firmer conclusions regarding the stellar metallicity gradient in M83.
\bigskip

We emphasize that in our comparison between stellar and nebular abundances we have not accounted for the effect of oxygen depletion onto interstellar dust grains, which in the interstellar medium in the solar neighbourhood is on the order of 0.1--0.2~dex (\citealt{Cartledge:2006, Jenkins:2009}). While the treatment of dust physics, including depletion, can be incorporated in photoionization models (\citealt{Groves:2004}), for a meaningful comparison with the stellar metallicities an upward correction to the gas-phase oxygen abundances due to dust depletion should be made when these have been derived via an empirical calibration or a direct measurement.
In the model grid by \citet{Kewley:2002}, used to calibrate the N2O2 method shown in panel b of Fig.~\ref{fig:hii},
the adopted oxygen depletion factor at solar metallicity is $-0.22$ dex, while more recently \citet{Dopita:2013} used $-0.07$ dex.
Empirical determinations in \hii\ regions by \citet{Mesa-Delgado:2009} and \citet{Peimbert:2010} provide 
depletion factors that are between $-0.08$ and $-0.12$~dex. The study of the Orion OB1 association by \citet{Simon-Diaz:2011} indicates a factor of approximately $-0.15$~dex. In the following discussion for simplicity we adopt a value of $-0.1$ dex.
Accounting for dust depletion would generally bring the nebular data shown in Fig.~\ref{fig:hii}   in better agreement with the stellar data in an absolute sense when considering empirically-calibrated abundance diagnostics, reducing the systematic discrepancy by $\sim$0.1 dex. Uncertainties in the amount of oxygen locked up in dust grains ultimately limit the precision with which we can compare gas-phase and stellar abundances. More optimistically, in the future we can also hope to learn about variations of dust depletion effects in different environments (\eg\ varying the metallicity) from this kind of comparisons.

%==========================================================================================================

\section{A chemical evolution model for M83}
In this section we introduce a chemical evolution model which reproduces the present-day spatial metallicity distribution over the entire disk of M83, as obtained both from 
blue supergiants and \hii\ regions. In order to apply the model, we require observed galactocentric radial profiles of the stellar and interstellar gas mass column densities, because the present-day 
metallicity reflects the continuous cycle of conversion of interstellar gas into stars and the recycling of nuclear processed stellar material back to the interstellar gas phase. 

In the inner disk
the ISM gas mass column density is dominated by molecular gas. We use the map of CO (1-0) emission at 115 GHz obtained by \citet{Crosthwaite:2002} with the NRAO 12m telescope at Kitt Peak and 
measure de-projected line intensities from a set of 10 arcsec-wide concentric tilted rings, which are then converted into azimuthally averaged $\rm H_2$ mass column densities, assuming the standard $X_{\rm CO}$ of $2\times10^{20}~\rm cm^{-2}~(K~km~s^{-1})^{-1}$, as described in 
\citet{Schruba:2011} and \citet{Kudritzki:2015}. For the neutral ISM gas component we re-analyze the map of \hi\ 21 cm line emission observed with the NRAO Very large Array (VLA) as part of the 
THINGS survey \citep{Walter:2008} and obtain line intensities (again from 10 arcsec-wide tilted rings), which are then turned into neutral hydrogen mass column densities as described in 
\citet{Kudritzki:2015}, following the prescriptions in \citet{Walter:2008}
 (see also \citealt{Leroy:2008} and \citealt{Bigiel:2010}). For the azimuthal averaging we follow \citet{Bigiel:2010}, who noted that in the strongly inhomogeneous filamentary
distribution of \hi\ in the outer disk ($R \ge 0.6$\,\rtf) only regions with a 
mass column density larger than 0.5\,\msun~pc$^{-2}$ correlate with the star formation activity. Thus, only pixels
with column densities larger than this value were taken into account in the azimuthal \hi\ average. To convert the ISM hydrogen masses into total gas masses, including helium and heavy metals, a multiplicative factor of 1.36 was applied.  

Stellar mass column densities are measured by surface photometry of mid-infrared images observed by the {\em Wide-field Infrared Survey Explorer} (WISE; \citealt{Wright:2010}, see also 
Sect.~\ref{sec:metals}) in the W1 band at 3.4$\mu$ and by Spitzer/IRAC at 3.6$\mu$. Again, the same set of tilted rings as for the ISM gas is used with a conversion of surface magnitudes into mass 
column densities as
described in \citet{Kudritzki:2015}. In this way a direct measurement of stellar mass is obtained in the range $0 \le R/R_{25} \le 1.15$. 
For the outer disk ($1.5 \le R/R_{25} \le 3.0$),
where the stellar mass column density is below the detection limit}, we use the observation by \citet{Bigiel:2010} that the star formation rate column density is closely correlated with the \hi\ mass column density via the proportionality 

\begin{equation}
\rm \Sigma_{SFR} = {1 \over \tau_{depl}}~ \Sigma_{H\,I}
\end{equation}

where $\rm \tau_{depl} \approx$  70 Gyr is the \hi\ depletion time. Assuming constant star formation as a function of time one can then approximate the present stellar mass column density in the outer disk by

\begin{equation}
\rm \Sigma_{\star} = {\tau_{disk} \over \tau_{depl}}~ \Sigma_{H\,I}.
\end{equation}

\begin{figure}
\center \includegraphics[width=1.01\columnwidth]{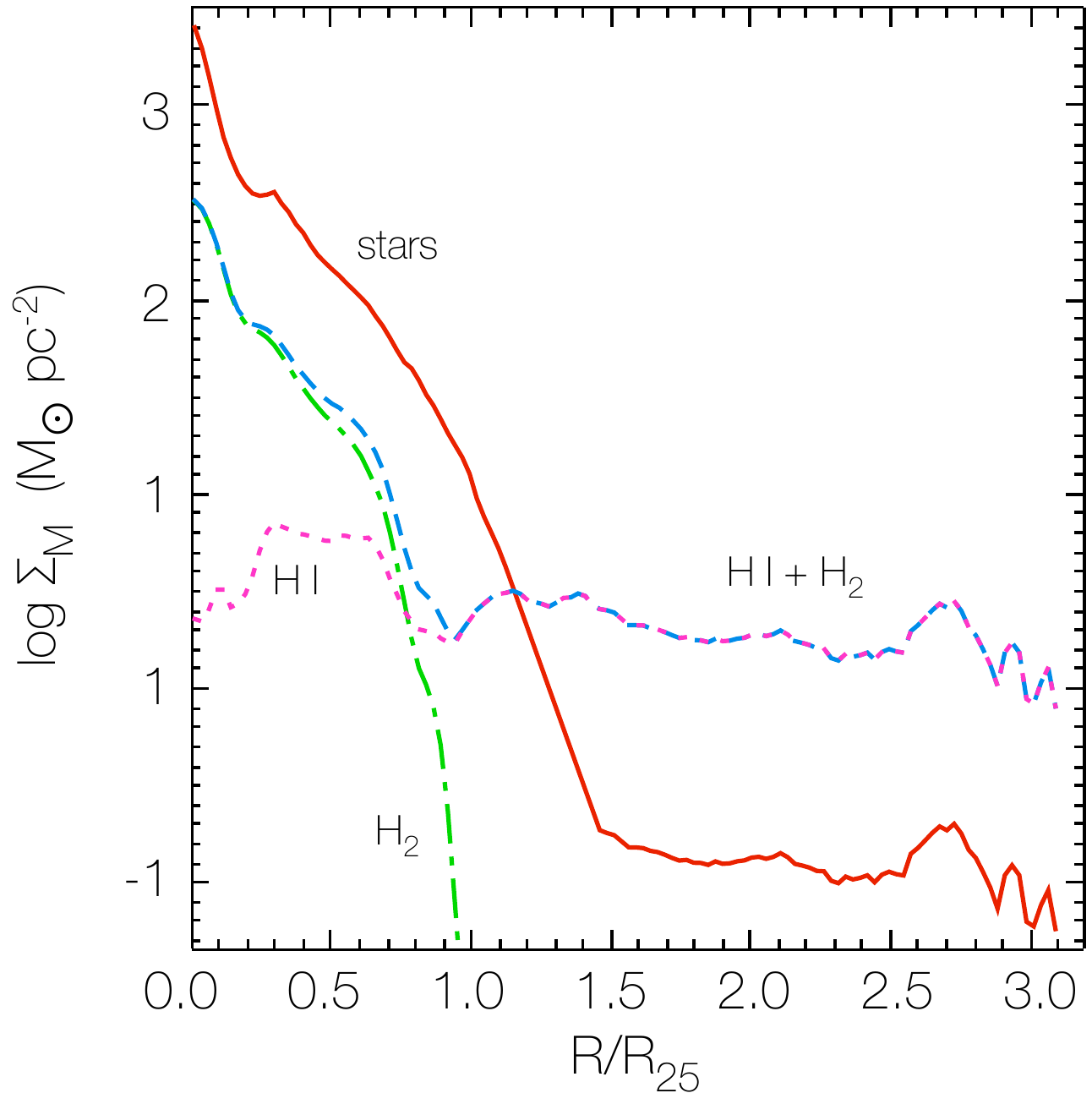}\medskip
\caption{Logarithm of M83 radial column density mass profiles for the different components used in the chemical evolution model. The gas masses have been corrected to account for the presence of helium and heavy metals.
}\label{fig:mass_profiles}
\end{figure}

\noindent
$\rm \tau_{disk}$ is the age of the outer disk, for which we assume 5 Gyr. 
For the transition region ($1.15 \le R/R_{25} \le 1.5$) we adopt an exponential decline with a scale length of 0.1078 in $R$/\rtf\ units, until the column density level of the outer disk is reached.
The resulting profiles of stellar, total gas, neutral and molecular gas masses are shown in Fig.~\ref{fig:mass_profiles}. 

For our modelling effort we use the chemical evolution model developed by \citet{Kudritzki:2015}. Contrary to the simple, closed box description, in which the evolution of the metallicity is tied solely to the ratio of stellar to gas mass (\citealt{Pagel:1975}), this model accounts for effects of gas inflows and outflows in regulating the radial  distribution of metals, and oxygen in particular (\eg\ \citealt{Edmunds:1990}). For simplicity, it assumes time- and location-invariant rates of mass infall $\Lambda = \dot{M}_{\rm accr}/\psi$ and mass outflow due to galactic winds $\eta = \dot{M}_{\rm loss}/\psi$ as a function of the star formation rate $\psi$. The observed galactic radial metallicity distribution is then described analytically as a function of the stellar and gas radial mass profiles, with the infall and outflow rates as free parameters, derived from comparing the observed metallicity radial profile with the model. 

\begin{figure}
\center \includegraphics[width=1.01\columnwidth]{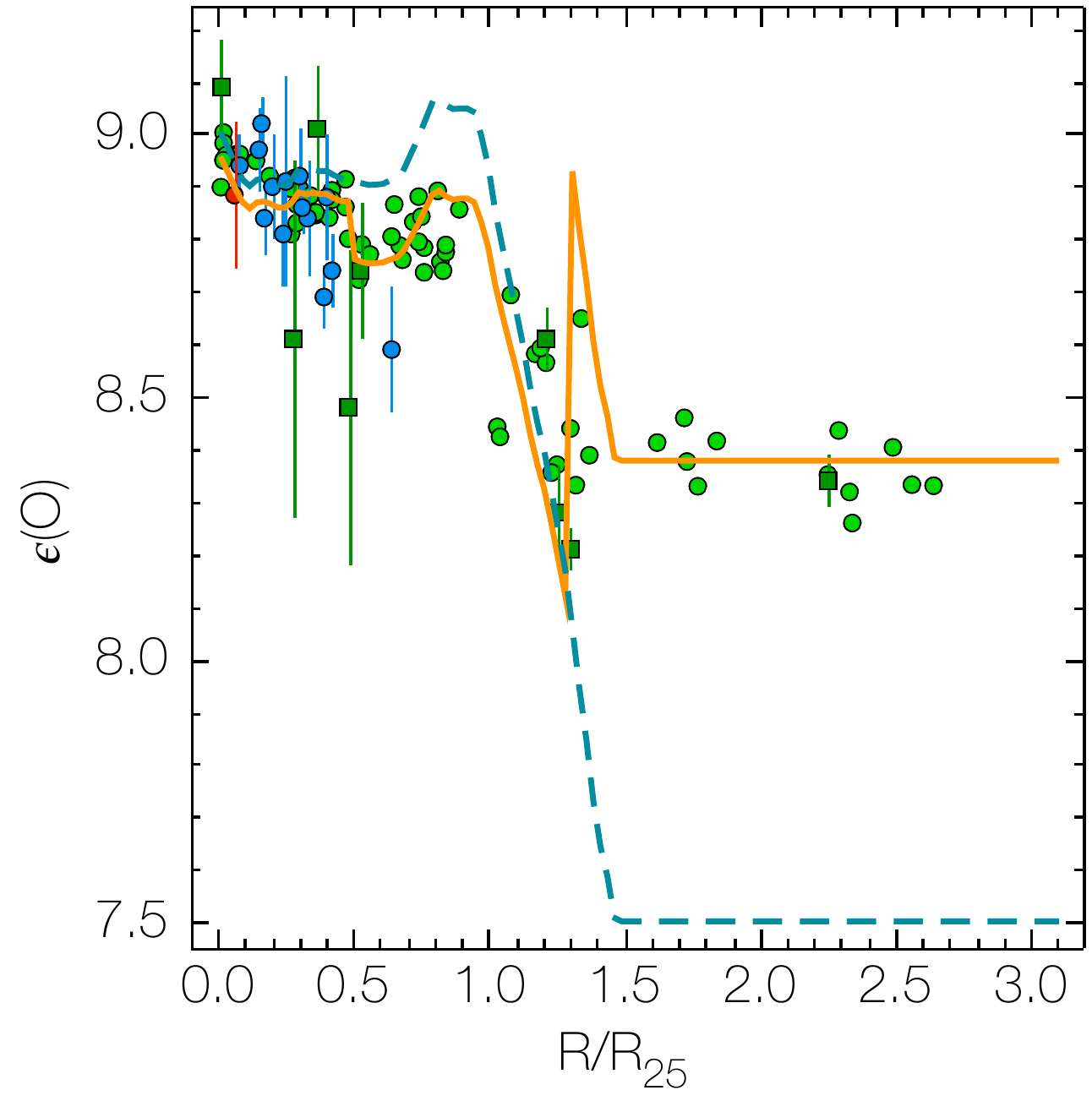}\medskip
\caption{{Comparison between the observed metallicities in M83 and our chemical evolution model. The data refer to the strong-line abundances obtained from the N2O2 diagnostic (B07 calibration, green dots), the auroral line-based abundances (green squares), the stellar metallicities (blue dots)  and the central super star cluster (red dot). Models: accounting for inner galactic winds and outer enriched infall (orange full line, see text) and closed-box model without outflow and infall (dashed cyan line).}}\label{fig:model}
\end{figure}

In a first step, we calculate a simple closed box model with $\eta = 0$ and $\Lambda = 0$, adopting the stellar and gas masses of Fig.~\ref{fig:mass_profiles} and the chemical yields and stellar mass return fractions of \citet{Kudritzki:2015}. The predicted oxygen abundance is shown in Fig.~\ref{fig:model} by the dashed cyan line.  The observations are displayed as green dots (strong-line abundances from the N2O2 B07 method, selected because of the small abundance scatter compared to other methods), green squares (auroral lines), blue dots (stellar metallicities) and a single red dot (the inner SSC). We adjusted the nebular N2O2 strong line abundances by $+0.2$~dex to account for the systematic difference found in Sect.~\ref{sec:sl}. The auroral abundances were shifted by $+0.1$~dex to compensate for the effects of dust depletion (see Sect.~\ref{sec:hii} and \ref{Sec:discussion}). In the inner region, $R/R_{25} \le 0.5$, the closed box model is only marginally off, predicting metallicities slightly too large when compared with most of the observed objects, while for $0.5 \le R/R_{25} \le 1.0$ the model metallicities are clearly too high.  For the outer disk the closed box model produces metallicities that are almost an order of magnitude too low.

The failure of the closed box model in the outer disk has already been pointed out by \citet{Bresolin:2012}, who postulated that the elevated levels of metal enrichment found in the extended UV disks of a few spiral galaxies, including M83, are consistent with an enriched infall scenario, in which metal-enriched gas inflows are responsible for the observed abundances, while the constant ratio of the  star formation rate to gas surface densities (\ie\ the star formation efficiency) would explain the flat outer gradient (see also \citealt{Kudritzki:2014} for the spiral galaxy NGC~3621). Our modeling now allows us to test these ideas, and to assign meaningful constraints to the metallicity of the infalling gas.

In the next step we apply an improved model which accounts for galactic wind outflows and accretion infall. We assume that the metallicity of the outflowing gas is equal to the actual metallicity of the ISM, whereas infall happens with a fixed metallicity, which could either be zero in case the galaxy accretes pristine gas from the cosmic web or larger than zero, if matter falls in from a halo enriched by mass outflow from the inner galactic disk. For the latter case we modify the analytical model by \citet{Kudritzki:2015} by replacing the yield $y_{\rm Z}$ in their equations (18), (21), (22) by $\tilde{y}_{\rm Z} = y_{\rm Z} + Z_{\rm infall}\, \Lambda$, where $Z_{\rm infall}$ is the metallicity mass fraction of the infalling gas.

We divide the radial range into three separate zones, in which we vary the mass outflow and inflow rates, and the metallicity of the infalling gas. In our modeling we find that the best-fitting solution requires no infall ($\Lambda = 0$) and moderate rates of outflow $\eta$ in the inner disk. The observations of the outer disk, on the other hand, require significant infall with gas which is already metal enriched to the level of a typical Local Group dwarf galaxy. Our adopted solution has the following parameters:

\smallskip

\begin{tabular}{l c c c c}
Zone	& 	$R$/\rtf\ 	& 	\eo$_{\rm infall}$		& 	$\Lambda$ 	&	$\eta$	\\[2pt]
 I   	&	0.0 -- 0.5  & 	0.00      				&	0.00		&	0.12	\\
II   	&	0.5 -- 1.3  &	0.00      				&	0.00		&	0.50	\\
III   	&	1.3 -- 3.0  &   8.20      				&	1.00		&	0.00	\\		
\end{tabular}
\medskip

%==========================================================================================================

%==========================================================================================================
Our model fit is displayed in Fig.~\ref{fig:model}. The model reproduces the spatial distribution of metallicity nicely. The case for an enrichment of the infalling gas, presumably coming from matter previously ejected by the inner regions of the disk, is evident. The O/H spike at $R=1.3$ \rtf\ is an artefact of the modeling procedure occurring at the beginning of zone III, where the ratio of stellar to gas mass still declines rapidly, while the metal enriched infall has already started. We also note that the model fit of zone III is not unique. As can be shown analytically from the modified equations in \citet{Kudritzki:2015}, every model with \eo$_{\rm infall} = 8.20 - \log(\Lambda) (\Lambda \neq 1)$ produces a similar fit. However, we can set an upper limit to $\Lambda$ for this degeneracy. For very high infall rates $\Lambda$ the chemical evolution model has solutions only for ratios of stellar to gas mass limited to

\begin{equation}
\rm \frac{\Sigma_{\star}}{\Sigma_{H\,I}} \le \frac{1}{\frac{5}{3}\Lambda - 1}
\end{equation}

(see \citealt{Kudritzki:2015}, Sect.~3, their case $\alpha \lneqq -1$). For the outer disk this limits the infall rate to 

\begin{equation}
\rm \Lambda \le \frac{3}{5} \left ( \frac{\tau_{depl}}{\tau_{disk}} + 1 \right ) = 9
\end{equation}

and the minimum metallicity of the infalling gas to \eo$_{\rm infall}$ = 7.25, comparable to an extremely metal-poor dwarf galaxy such as I~Zw~18.

Independent of this degeneracy, the chemical evolution modeling procedure we carried out demonstrates two important requirements to reproduce the abundances in the outer disk of M83: a chemically enriched gas inflow and a star formation efficiency that is roughly constant with radius, in line with the suggestions made by \citet{Bresolin:2012}. It also indicates that a linear fit to the observations with a single gradient in the inner region might be too simple an approach to capture the underlying physics responsible for the spatial distribution of metallicity.

%==========================================================================================================

%==========================================================================================================

%==========================================================================================================
\section{Discussion}\label{Sec:discussion}
\subsection{Strong-line methods}

The comparison we carried out in Sect.~\ref{sec:hii} reveals that most of the nebular diagnostics we considered yield abundances that do not agree with the metallicities of the blue supergiant stars in M83. This appears to be true for both empirically- and theoretically-calibrated diagnostics.
The potential perils of systematic uncertainties, although difficult to estimate, should be kept in mind. For example, abundance offsets could result from a significant mismatch in physical properties between the nebulae in M83 and the calibrating samples or models used for the abundance diagnostics. In this regard, we do note that the nebular N/O ratio in M83 appears to be higher than average (\citealt{Bresolin:2005}), albeit the uncertainties are large, and this could affect the abundances derived from diagnostics involving the nitrogen lines. However, a higher N/O ratio would lead to overestimate the nebular O/H ratio (\citealt{Perez-Montero:2009a}), opposite to what the comparison with the stellar metallicities suggests. For systematics concerning stellar abundances, we refer to \citet{Przybilla:2006} and \citet{Nieva:2012} and references therein.

Panels a--d in Fig.~\ref{fig:hii} indicate that at the highest metallicities considered in this work on M83 (nearly 2$\times$ solar) some of the theoretical calibrations can produce nebular abundances in good agreement with the stellar metallicities we measured, in particular, the O3N2 method (panel c), whose calibration by \citet{Pettini:2004} at the high-metallicity end relies on photoionization models
(this holds also after adding 0.1 dex to the nebular abundances to account for dust depletion on dust grains).
On the other hand, panels e--h  show that empirical, \te-based calibrations of strong-line methods
 yield results that, approximately above the solar O/H value, lie $\sim$0.2~dex below the stellar metallicities. At the same time, some of the auroral line-based nebular abundances appear to agree with the stellar metallicities even very close to the center of M83, where the metallicity is highest.
\medskip

At face value, and considering the blue supergiant surface chemical abundances to be representative of the `true' metallicity of the young populations of M83, these results suggest the existence of a problem with the empirical calibrations, \ie\ that they progressively underestimate O/H with increasing metallicity, by $\sim$0.1--0.2 dex around 2$\times$ the solar value (correcting for 0.1 dex due to dust depletion).

If in the following we assume this to be correct, this could result from the well-known difficulty for the empirical methods to establish the calibrating samples of high-metallicity \hii\ regions, which rely on the detection of faint auroral lines and somewhat uncertain relationships used to infer, for example, the temperature of the \oiii-emitting nebular zone from the temperature measured for the \siii- or \nii-emitting zones (\eg\ \citealt{Garnett:1992}). 
It is thus possible that the empirical calibrations are affected by a selection bias, whereby the \hii\ regions with the strongest auroral lines (corresponding to higher gas electron temperatures and lower metallicities) are preferentially measured at high oxygen abundances. While a few high-metallicity \hii\ regions could still be providing reliable abundances, as seen also in the case of M83, more generally the calibrating samples could be biased to low abundances.  
A completely different interpretation is that we might be detecting the bias  predicted 
by \citet{Stasinska:2005} to occur due to \hii\ region temperature stratification. According to this work, the direct method could underestimate the abundance by 0.2~dex or more above the solar value. 
Nebular abundances that are systematically higher than those derived from the direct method are also obtained from the use of recombination lines (as formalized by the presence of an abundance discrepancy factor ADF, \citealt{Garcia-Rojas:2007}), by an amount that is comparable to the difference we observe in the central regions of M83.
The most popular interpretation for this discrepancy is given in terms of temperature fluctuations (\citealt{Peimbert:1967, Peimbert:2013}), but other explanations have also been proposed, such as the presence of metal-rich inclusions (\citealt{Tsamis:2003, Stasinska:2007a}). Alternatively, deviations from the thermal electron velocity distribution commonly assumed for ionized nebulae have been invoked  (\citealt{Nicholls:2012, Nicholls:2013}).
\medskip

\subsubsection{A recommended strong-line method?} 
It bears on the initial motivation of our work to try and identify which, if any, of the strong-line methods we looked at can be recommended for extragalactic emission-line abundance studies in order to obtain metallicities that are in agreement, in an absolute sense, with current and published results based on stellar spectroscopy. We emphasize again that such an approach is encouraged by the relatively small systematic uncertainties in the stellar abundances, and the good agreement for the metallicities determined independently for massive hot and cool stars, from analyses carried out in different wavelength regimes (\citealt{Gazak:2014, Gazak:2015, Davies:2015}), which boosts our confidence on the metallicity scale defined by massive stars.

From our discussion in Sect.~\ref{sec:hii} the O3N2 diagnostic calibrated by \citet{Pettini:2004} stands out as the only one providing \hii\ region abundances that are consistent with our stellar metallicities in M83, which are all but one above the solar value. In the similarly high metallicity (\eo\ $>$ 8.6) environment of the galaxy M81 we reach the same conclusion, analyzing the supergiant data from \citet{Kudritzki:2012} and the nebular emission fluxes from \citet{Patterson:2012} and \citet{Arellano-Cordova:2016}. Keeping in mind the statistical nature of strong-line diagnostics (\ie\ the fact that they can fail on individual objects) we can extend this statement to include lower metallicities by looking, for example, at our study of NGC~300 (\citealt{Bresolin:2009a}). We find that in this case (\eo\ $<$ 8.6) the radial trend of the stellar metallicities is equally well reproduced by O3N2 (PP4), the ONS and the $R$ methods, if a modest dust depletion factor is introduced. In summary, the use of O3N2 (PP4) for extragalactic \hii\ regions provides \eo\ values that are consistent with the metallicity scale defined by our stellar work across a wide metallicity range, 
8.1 $\lesssim$ \eo\ $\lesssim$ 9.

%==========================================================================================================
\floattable
\begin{deluxetable}{lccccccccc}

\tablecolumns{10}
\tablewidth{0pt}
\tablecaption{Abundance data for objects with stellar and nebular abundance information.\label{table:celrl}}

\tablehead{
\colhead{Object}	     							&
\multicolumn{3}{c}{$\rm\epsilon(O)$: R\,=\,0}			&
\multicolumn{3}{c}{$\rm\epsilon(O)$: R\,=\,0.4~\rtf}	&
\multicolumn{3}{c}{References}\\[0.5mm]
\colhead{}       			&
\colhead{stars}       		&
\multicolumn{2}{c}{\hii\ regions}	&
\colhead{stars}       		&
\multicolumn{2}{c}{\hii\ regions}	&
\colhead{stars}  			&
\colhead{\cel}				&
\colhead{\rl}\\[0.5mm]	
\colhead{}       			&
\colhead{}       		&
\colhead{\cel}     			&
\colhead{\rl}      			&
\colhead{}       		&
\colhead{\cel}     			&
\colhead{\rl}      			&
\colhead{}					&
\colhead{}					&
\colhead{}  } 					
\startdata
\\[-4mm]
Sextans A	&  $7.70 \pm 0.07$	&	$7.49 \pm 0.06$	& \nodata 			& 	\nodata & \nodata & \nodata 	&	K04	&	K05		& \\ 
WLM			&  $7.82 \pm 0.06$	&	$7.82 \pm 0.09$	& \nodata 			& 	\nodata & \nodata & \nodata 	&	U08	&	L05		&  \\  
IC~1613		&  $7.90 \pm 0.08$	&	$7.78 \pm 0.07$	& \nodata 			& 	\nodata & \nodata & \nodata 	&	B07	&	B07		&  \\  
NGC~3109	&  $8.02 \pm 0.13$	&	$7.81 \pm 0.08$	& \nodata 			& 	\nodata & \nodata & \nodata 	&	H14	&	P07		&  \\  
~~~~~''		&  $7.76 \pm 0.07$	&	$7.81 \pm 0.08$	& \nodata 			& 	\nodata & \nodata & \nodata 	&	E07	&	P07		&  \\  
NGC~6822	&  $8.08 \pm 0.21$	&	$8.14 \pm 0.08$	& $8.37 \pm 0.09$ 	& 	\nodata & \nodata & \nodata 	&	P15	&	L06		& P05  \\  
SMC			&  $8.06 \pm 0.10$	&	$8.05 \pm 0.09$	& $8.24 \pm 0.16$	& 	\nodata & \nodata & \nodata 	&	H07	&	B07		& PG12  \\  
LMC			&  $8.33 \pm 0.08$	&	$8.40 \pm 0.10$	& $8.54 \pm 0.05$	& 	\nodata & \nodata & \nodata 	&	H07	&	B07		& P03  \\  
NGC~55		&  $8.32 \pm 0.06$	&	$8.21 \pm 0.10$	& \nodata			& 	\nodata & \nodata & \nodata 	&	K16	&	T03		&  \\  
NGC~300		&  $8.59 \pm 0.05$	&	$8.59 \pm 0.02$	& $8.71 \pm 0.10$	& 	$8.42 \pm 0.06$ & $8.43 \pm 0.02$ & $8.65 \pm 0.12$ 	&	K08	&	B09		& T16 \\  
M33			&  $8.78 \pm 0.04$	&	$8.51 \pm 0.04$	& $8.76 \pm 0.07$	& 	$8.49 \pm 0.05$ & $8.36 \pm 0.05$ & $8.63 \pm 0.09$ 	&	U09	&	B11		& T16 \\  
M31			&  $8.99 \pm 0.10$	&	$8.74 \pm 0.20$	& $8.94 \pm 0.03$	& 	$8.74 \pm 0.10$ & $8.51 \pm 0.21$ & $8.69 \pm 0.03$ 	&	Z12	&	Z12		& E09 \\  
%M81		&  $8.98 \pm 0.06$	&	$8.26 \pm 0.10$	& \nodata			& 	$8.81 \pm 0.07$ & $8.25 \pm 0.12$ & \nodata			 	&	K12	&	A16		&  \\  
M81			&  $8.98 \pm 0.06$	&	$8.86 \pm 0.13$	& \nodata			& 	$8.81 \pm 0.07$ & $8.72 \pm 0.13$ & \nodata			 	&	K12	&	P12		&  \\  
M42			&  $8.74 \pm 0.04$	&	$8.53 \pm 0.01$	& $8.65 \pm 0.03$	& 	\nodata & \nodata & \nodata 	&	S11	&	E04		& S11  \\  
M83			&  $9.04 \pm 0.04$	&	$8.90 \pm 0.19$	& \nodata			& 	$8.78 \pm 0.07$ & $8.73 \pm 0.27$ & \nodata			 	&	This	&	 work		&  \\ [1mm] 
\enddata
\tablerefs{{\sc Stars:} K04: \citet{Kaufer:2004}; U08: \citet{Urbaneja:2008}; B07: \citet{Bresolin:2007a}; H14: \citet{Hosek:2014}; E07: \citet{Evans:2007}; P15: \citet{Patrick:2015}; H07: \citet{Hunter:2007}; K16: \citet{Kudritzki:2016}; K08: \citet{Kudritzki:2008}; U09: \citet{U:2009}; Z12: \citet{Zurita:2012}; K12: \citet{Kudritzki:2012}; S11: \citet{Simon-Diaz:2011}.
~---{\cel:} K05: \citet{Kniazev:2005}; L05: \citet{Lee:2005}; B07: \citet{Bresolin:2007a}; P07: \citet{Pena:2007}; L06: \citet{Lee:2006}; T03: \citet{Tullmann:2003}; B09: \citet{Bresolin:2009a}
B11: \citet{Bresolin:2011a}; Z12: \citet{Zurita:2012}; P12: \citet{Patterson:2012}; E04: \citet{Esteban:2004}.
~---{\rl:} P05: \citet{Peimbert:2005}; PG12: \citet{Pena-Guerrero:2012}; P03: \citet{Peimbert:2003}; T16: \citet{Toribio-San-Cipriano:2016}; E09: \citet{Esteban:2009}; S11: \citet{Simon-Diaz:2011}.}
\tablecomments{All \cel-based abundances redetermined with consistent and updated atomic data (see text).}
\end{deluxetable}
%==========================================================================================================
\subsection{Stellar \vs\ nebular abundances: auroral and recombination lines}
Despite the complexity of the physics of ionized nebulae, which hinders the resolution of issues related to their temperature and density structure,
and in view of the urgency to understand how to select the correct absolute abundance  scale,
it is worthwile to test empirically whether the difference between stellar and nebular direct abundances remains constant with metallicity, as is the case for the difference obtained using \cel s and \rl s ($\sim$0.2 dex, \citealt{Garcia-Rojas:2007}). For this purpose, we have assembled published data on stellar abundances for young stars and \hii\ regions in nearby galaxies and the Milky Way, as summarized in Table~\ref{table:celrl}. The nebular oxygen abundances refer to \cel-based determinations and, for seven objects, \rl-based results. The latter refer mostly to single \hii\ regions in different galaxies, while \cel\ measurements are typically available for several \hii\ regions. For irregular galaxies, due to their spatially homogeneous abundance distribution or their flat/very shallow metallicity gradients, we report mean abundance values, while for spirals we use the available radial gradient information to obtain the metallicity both at the center and at 0.4~\rtf. For several of the galaxies reported in Table~\ref{table:celrl} we used the data compilation from \citet{Bresolin:2011}, who re-analyzed published emission line fluxes in order to homogenize the derived abundances, using a set of atomic data consistent with the work on NGC~300 by \citet{Bresolin:2009a}. For the present work we re-determined all the \te-based abundances using IRAF's {\em nebular} package, with the atomic parameters used in \citet[Table~5]{Bresolin:2009a} but updating the \ion{O}{3} collision strengths from \citet{Palay:2012}, and re-deriving radial gradients when necessary. The updated \ion{O}{3} collision strengths  determined an increase in \eo\ of typically 0.02--0.04~dex.
It is worth pointing out that our comparison is mostly of a statistical nature, because the ideal situation in which stellar and nebular abundances are simultaneously available for young stars and their parent gas cloud, as in the case of the Orion nebula in the Milky Way, is still not realized with current data in extragalactic systems.

%==========================================================================================================

\begin{figure*}
\center \includegraphics[width=1.7\columnwidth]{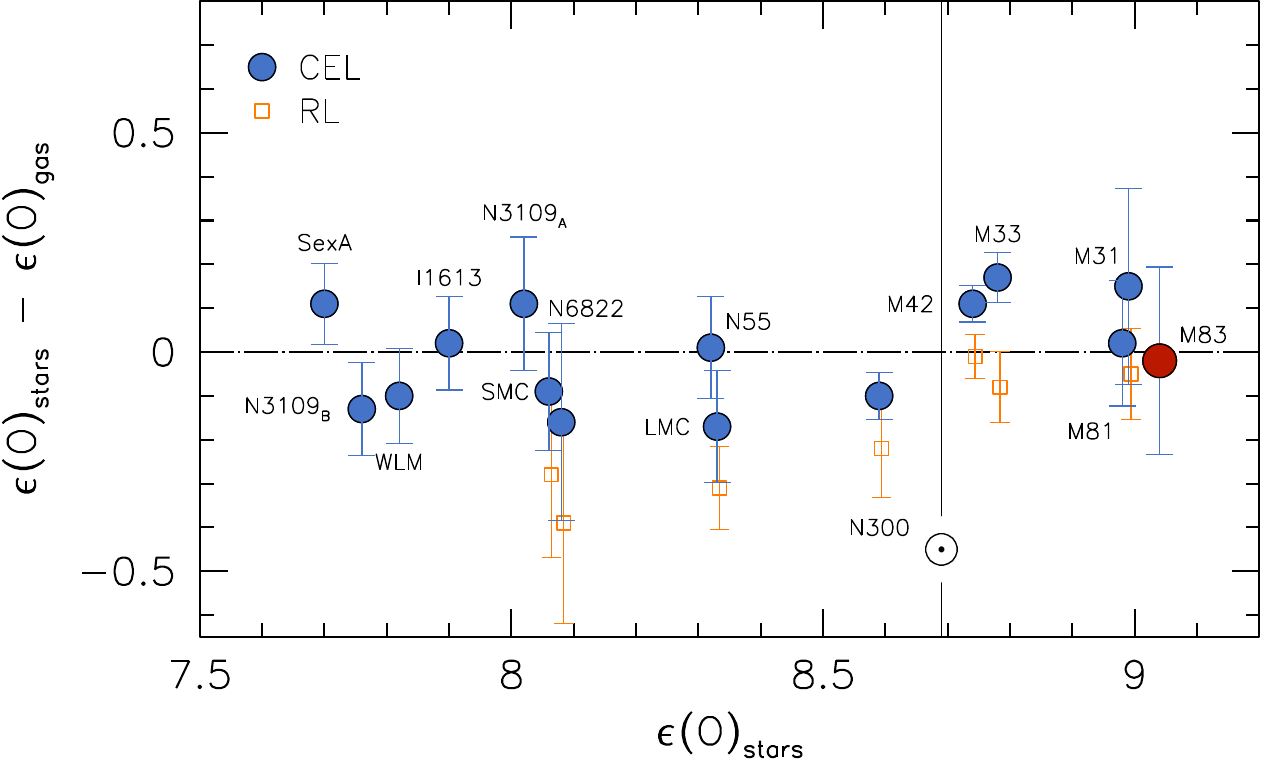}\medskip
\caption{Difference in metallicity between young stars and ionized gas for a sample extracted from the literature and the M83 data presented here. We have added 0.1 dex to the gas metallicities reported in Table~\ref{table:celrl} to account for dust depletion. For spiral galaxies the metallicities correspond to the central values. We use blue circles and orange squares for nebular oxygen abundances determined from the direct method and from recombination lines, respectively. The adopted solar O/H value is shown by the vertical line.}\label{fig:hiistars}
\end{figure*}

%==========================================================================================================

\medskip
In Fig.~\ref{fig:hiistars} we show the difference between stellar and nebular abundances as a function of stellar metallicity. We added 0.1~dex to the \hii\ region abundances included in Table~\ref{table:celrl} to account for the effect of depletion onto dust grains.  
For spiral galaxies we use the central metallicity values (our main conclusions do not change if we use the characteristic metallicity at 0.4~\rtf). %In Appendix~A we display the same diagram, using metallicities measured from the known abundance gradients at 0.4~\rtf\ for spiral galaxies. 
The blue dots refer to the quantity $\rm \Delta\epsilon(O)_{CEL}$, the 
(stars$-$gas) metallicity difference, using direct abundances for \hii\ regions. The orange open square symbols are used for the corresponding quantity $\rm \Delta\epsilon(O)_{RL}$ using the nebular \rl s instead to estimate the gaseous abundances. In order to support our interpretation,
we comment on the following objects:

\let\origdescription\description
\renewenvironment{description}{
  \setlength{\leftmargini}{0em}
  \origdescription
  \setlength{\itemindent}{0em}
}
%{\endlist}

\begin{description}

\item[\rm Sextans~A] The spectral data we used for the nebular abundance of three \hii\ regions, from \citet{Kniazev:2005}, do not cover the \oii\lin3727 line, and the resulting $\rm O^+/H^+$ abundance relies on the \oii\llin7320--7330 auroral lines instead, and as such we suspect that it is  subject to a higher level of uncertainty than reported (see \citealt{Kennicutt:2003}).

\item[\rm NGC~3109] There is a discrepancy between the metallicities of B- and A-type supergiants from \citet{Evans:2007} and \citet{Hosek:2014}, respectively. We use both measurements in Fig.~\ref{fig:hiistars}, using the stellar type (B or A) as a subscript to the galaxy name.

\item[\rm NGC~6822] We use the mean metallicity of the 11 red supergiants studied by \citet{Patrick:2015}, with a $-0.086$~dex correction to account for the difference in the adopted solar metallicity value (see Sect.~\ref{sec:hii} with respect to the MARCS model atmospheres used for red supergiants). Although we are not using blue supergiants for this galaxy, we point out that red supergiants have been shown by \citet{Gazak:2015} to provide chemical abundances that are in excellent agreement with blue supergiants.

\item[\rm SMC] We use the \rl\ measurements from \citet{Pena-Guerrero:2012} for the two \hii\ regions NGC~456 and NGC~460, taking the weighted average of the published, gas-phase results. We do not include the study of N66 by \citet{Tsamis:2003}, which is highly discrepant relative to the stellar and \cel-based metallicities, with \eo\,=\,8.47, but without an estimate of the uncertainty.

\item[\rm M31] The abundance gradient in the Andromeda Galaxy is still quite uncertain. For the estimation of the quantities in Table~\ref{table:celrl} we relied on the gradient determined from \cel\ by \citet{Zurita:2012}, and used the same slope to estimate the values for \rl s and stars. Based on \citet{Zurita:2012} and \citet{Esteban:2009} we used a value $\rm \Delta\epsilon(O)$ relative to the \cel s of +0.25~dex and +0.2~dex for stars and \rl s, respectively. 

%\item[\rm 81] We use the nebular emission lines from \citet{Patterson:2012}. Recently \citet{Arellano-Cordova:2016} presented a new direct abundance analysis in this galay, obtaining much lower values for O/H than \citet{Patterson:2012}. Considering these  lower \eo\ values for the \hii\ regions would give $\rm \Delta\epsilon(O)_{CEL} \simeq $ value

\item[\rm M42] We include data for the Orion nebula and the Orion OB1 stellar association in the Milky Way. The abundance results for this object  are consistent with other measurements of the chemical abundances in the local neighbourhood (\eg\ \citealt{Nieva:2012, Garcia-Rojas:2014}), not included in the figure for clarity. We re-derived the \cel-based nebular oxygen abundance using the data by \citet{Esteban:2004}, and following the same procedure as in \citet{Simon-Diaz:2011}, \ie\ using the \nii\ temperature for the $\rm O^+$ region, and the electron density from the \oii\ 3726/3727~\AA\ line ratio. The effect of the updated \ion{O}{3} collision strengths
on the final oxygen abundance is minor ($\sim 0.01$~dex).

\item[\rm M83] As we mentioned earlier, the auroral line-based gradient, that we used to estimate the central abundance, is quite uncertain. Nevertheless, the central abundance that we adopt is close to the value we measure for the central \hii\ region.

\end{description}
\bigskip

Focusing on $\rm \Delta\epsilon(O)_{CEL}$ first, we note that this quantity appears to be largely independent of metallicity. %, although above the solar abundance no negative value is observed. 
Fig.~\ref{fig:hiistars} suggests that the direct method yields metallicities that could lie, on average, below the stellar ones at high metallicity, but does not seem to be true for all objects. We divided (arbitrarily) the sample at \eo\,=\,8.7 and performed a weighted mean for different metallicity ranges, as summarized below:
\smallskip

\begin{tabular}{l c}
Range					& $\rm \Delta\epsilon(O)_{CEL}$ -- Weighted mean \\[2pt]
$\rm \epsilon(O) < 8.7$ & $-0.05 \pm 0.09$ \\
$\rm \epsilon(O) > 8.7$ & $+0.12 \pm 0.04$ \\
All						& $+0.03 \pm 0.11$ \\
\end{tabular}
\medskip

The difference between high and low metallicity is marginally significant ($\sim 1 \sigma$). The point remains that for some objects with small observational errors (M33, M42 and other Galactic objects not included in Fig.~\ref{fig:hiistars}, \eg\ the Cocoon Nebula, \citealt{Garcia-Rojas:2014}) the direct method underestimates the stellar metallicity by $\sim$0.1~dex, even considering the dust depletion correction. %At metallicities below solar, on the other hand, by and large the stellar and the dust-corrected direct metallicities agree to within ~0.1~dex, with apparently no systematic offset.

Turning to $\rm \Delta\epsilon(O)_{RL}$, as shown by the seven open square symbols in Fig.~\ref{fig:hiistars}, we notice a somewhat opposite behavior. The agreement with the stellar metallicities is excellent in the high-abundance regime, a result that has been pointed out already by several authors (\eg\ \citealt{Simon-Diaz:2011}). At lower metallicities, however, the \rl-based nebular abundances tend to diverge from the stellar ones. The mean offset for the four data points at \eo\,$<8.7$ is $-0.28 \pm 0.05$, after the 0.1~dex correction for dust depletion.
To our knowledge, this is the first time that this effect has been identified or emphasized.
%This discrepancy depends only minimally on our choice of stellar abundance indicators and stellar types. 
We examine here briefly the four data points in Fig.~\ref{fig:hiistars} that indicate a significant difference between stellar and \rl-based metallicities.\\

\noindent
SMC and LMC: The stellar metallicities and mean \eo\ values of the Small and Magellanic Clouds are known to quite good precision from the VLT-FLAMES survey (\citealt{Hunter:2007}), in which the chemical  abundances of B-type stars are obtained with the same non-LTE {\sc fastwind} code (\citealt{Puls:2005}) utilized for other objects included in Fig.~\ref{fig:hiistars} (\eg\ M42, NGC~300, WLM, NGC~3109), which ensures some level of homogeneity in our analysis. 
We also note that for the LMC the \citet{Hunter:2007} metallicity agrees very well with the most recent study of 90 blue supergiants
by \citet{Urbaneja:2016}.
The \rl s have been studied in the two SMC nebulae mentioned earlier and in 30~Dor for the LMC.
\\

\noindent
NGC~300: \citet{Bresolin:2009a} found very good agreement between the absolute abundances determined from A and B supergiants, which rely upon different diagnostic lines as well as stellar models. Moreover, \citet{Urbaneja:2016} demonstrated the absence of systematic effects when the spectral analysis is carried out from spectra of high (as used in the LMC/SMC) or medium (as used in NGC~300) resolution.
The $\rm \Delta\epsilon(O)_{RL}$ value we used for this galaxy does not depend on the use of central abundances only, as can be seen from the work on the metal \rl s by \citet{Toribio-San-Cipriano:2016}\\

\noindent
NGC~6822: We have used the recent metallicities for 11 red supergiants  from \citet{Patrick:2015}, which are in good agreement 
with the overall metallicity obtained from B-type supergiants by \citet{Muschielok:1999} and from 2 A-type supergiants by \citet{Venn:2001}.\\
\smallskip

We note that the mean difference between \rl- and \cel-based abundances is 0.16\,$\pm$\,0.05 for the seven objects included in Fig.~\ref{fig:hiistars}, consistent with the value for the oxygen  ADF\,=\,0.26\,$\pm$\,0.09 measured by \citet{Esteban:2009} for a sample of extragalactic \hii\ regions and with other determinations in the Milky Way (\eg\ \citealt{Garcia-Rojas:2007}).
\smallskip

An in-depth discussion of our results within the context of the non-equilibrium $\kappa$ electron energy distribution lies outside the scopes of this paper. However, it is worth recalling that the assumption of a 
$\kappa$ distribution has a profound impact on the abundances derived from \cel s, due to the strong sensitivity of these lines to the gas temperature (see \citealt{Nicholls:2012, Nicholls:2013} for details).
In fact, the assumption of even a moderate deviation from the Maxwellian energy distribution can explain the 
ADF observed in Galactic and extragalactic \hii\ regions, and similarly the abundance offset between theoretically-calibrated strong line abundance determination methods and the direct method.
We do note however that the photoionization models presented by \citet[see their Fig.~32]{Dopita:2013}, calculated for $\kappa = 20$, predict that this offset, which is roughly constant with metallicity below the solar value, increases rapidly for higher metallicities. \citet[Fig.~9]{Blanc:2015} also illustrated a difference between \rl\ abundances and those derived from  photoionization models that increases with metallicity. We suggest that this effect, that appears to be on the order of 0.2 dex, mirrors the behavior of $\rm \Delta\epsilon(O)_{RL}$ seen in Fig.~\ref{fig:hiistars}.

%==========================================================================================================
\section{Summary}
In this paper we have highlighted the importance of carrying out stellar spectroscopy of individual massive stars in nearby galaxies as a means to test the poorly understood systematic uncertainties of present-day nebular abundance diagnostics currently in use. This approach appears to be particularly relevant in a high metallicity, super-solar galactic environment, as encountered in the relatively nearby galaxy M83, because abundance biases that can affect the direct method should be more easily detected.

Within the context of a long-term program based on the quantitative stellar spectroscopy of blue supergiant stars in nearby galaxies, we have measured stellar parameters and metallicities for 14 A-type supergiants in the inner disk of M83. 
We have derived a spectroscopic distance to M83, based on the flux-weighted--luminosity relationship, finding an excellent agreement with alternative extragalactic distance determination methods.
We have used the metallicity information to provide a new data point in a version of the local galaxy mass-metallicity relation that avoids the use of \hii\ region emission line data, and discussed how this can be useful for an independent test of the shape and zero-point of the relation itself. We presented a chemical evolution model, tailored to reproduce the radial abundance gradient of this galaxy out to almost 3\,\rtf, that is able to quantify the metallicity of the gas infalling into the outer regions, and that is responsible for the chemical enrichment of the outer disk, as observed by \citet{Bresolin:2009}.

We then focused on the comparative analysis of present-day metallicities in M83, from measurements based on \hii\ regions (using the direct method and six different strong line diagnostics) and blue supergiants. We found that
\te-based abundances determined in the inner disk of M83 are in relatively good agreement with the stellar metallicities, once a $\sim$0.1 dex correction to the nebular oxygen abundance due to dust depletion is accounted for. However, around the solar metallicity and above oxygen abundances estimated from most
strong line methods calibrated empirically from \hii\ regions where the direct method can be applied tend to underestimate the stellar abundances. We argue that this can be related to difficulties in selecting the appropriate calibration samples at high metallicity. We find that among existing strong-line methods, O3N2 as calibrated by \citet{Pettini:2004} gives nebular abundances that are in best agreement with the stellar metallicities when radial abundance gradients are analyzed.

We confirm that metal recombination lines are in excellent agreement with stellar abundances for high metallicity systems (\eg\ the Orion nebula), but provide evidence that in more metal-poor environments they tend to underestimate the stellar metallicities by a significant amount, while the direct method does not seem to be systematically offset from the stars, except at high metallicity.
Future observations of \rl-based abundances in selected low-metallicity galaxies would be helpful to shed light on this point.

%==========================================================================================================
\acknowledgments
FB and RPK acknowledge support by the National Science Foundation under grants AST-1008789 and AST-1108906. RPK, MAU, WG were supported by the Munich Institute for Astro- and Particle Physics (MIAPP) of the DFG cluster of excellence ''Origin and Structure of the Universe''. WG and GP gratefully acknowledge financial support for this work received from the BASAL Centro de Astrof\'isica y Tecnolog\'ias Afines (CATA), PFB-06/2007. WG also acknowledges support from the Millenium Institute of Astrophysics (MAS) of the Iniciativa Cient\'ifica Milenio del Ministerio de Econom\'ia, Fomento y Turismo de Chile, grant IC120009. In addition, support from the Ideas Plus grant of the Polish Ministry of Science and Higher Education and TEAM subsidies of the Foundation for Polish Science (FNP) is acknowledged by GP. 
This work is based on observations collected at the European Organisation for Astronomical Research in the Southern Hemisphere under ESO programmes 091.D-0147(B) and 095.D-0021(B).\\

\vspace{5mm}
\facilities{VLT:Antu (FORS2), HST (WFC3)}

\software{PyRAF, SM, IDL, DOLPHOT, APLpy}

%==========================================================================================================

%\begin{thebibliography}{}

\bibliographystyle{aasjournal}
\bibliography{/Users/fabio/PDF-Papers/Papers}

%\end{thebibliography}

\end{document}